\documentclass{article}
\usepackage{a4,amssymb, amsmath, amsthm,dsfont, amsfonts}
\usepackage[a4paper, total={6in, 9in}]{geometry}
\usepackage{color}
\usepackage[utf8]{inputenc}
\usepackage[T1]{fontenc}
\usepackage{mathtools}
\usepackage{nicefrac}
\usepackage{csquotes}
\usepackage{empheq}
\usepackage{bbm}
\usepackage{graphicx}
\usepackage[normalem]{ulem}

\makeatletter
\newcommand*{\rom}[1]{\expandafter\@slowromancap\romannumeral #1@}
\makeatother

\usepackage{ upgreek }
 
\usepackage{amsmath,amssymb,amsthm,mathrsfs,amsfonts}

\newtheorem{thm}{Theorem}[section]

\theoremstyle{remark}
\newtheorem{rem}[thm]{Remark}

\usepackage[square,sort,comma,numbers]{natbib}

\usepackage{amsmath}
\newcommand*\diff{\mathop{}\!\mathrm{d}}

\newcommand\red[1]{\color{red}#1} 

\usepackage{wrapfig}

\usepackage[unicode=true,
 bookmarks=true,bookmarksnumbered=true,bookmarksopen=true,bookmarksopenlevel=1,
 breaklinks=false,pdfborder={0 0 0},backref=false,colorlinks=true]
 {hyperref}
\hypersetup{
 linkcolor=blue, urlcolor=marineblue, citecolor=blue, pdfstartview={FitH}, unicode=true}

\theoremstyle{definition}

\usepackage{multicol}
\usepackage{blindtext}

\newcommand {\tp} { {\tilde{\psi}} }
\newcommand {\tph} { {\tilde{\phi}} }

\newcommand{\e}{\varepsilon}

\usepackage[toc,page]{appendix}

\usepackage{graphicx,epstopdf}
\ifpdf
  \DeclareGraphicsExtensions{.eps,.pdf,.png,.jpg}
\else
  \DeclareGraphicsExtensions{.eps}
\fi

\usepackage{subfig}

\title{Motility switching and front-back synchronisation in polarized cells}

\author{Gissell Estrada-Rodriguez\thanks{Basque Center for Applied Mathematics, 48009 Bilbo, Bizkaia, Spain. email: {estradarodriguez@ljll.math.upmc.fr}}
\and Benoit Perthame\thanks{Sorbonne Universit\'e, Universit\'e de Paris, CNRS, Inria, Laboratoire Jacques-Louis Lions, F-75005 Paris, France. email: {Benoit.Perthame@sorbonne-universite.fr}
}
}

\date{}

\begin{document}

\maketitle

\begin{abstract}
    The combination of protrusions and retractions in the movement of polarized cells  leads to understand the effect of possible synchronisation between the two ends of the cells. This synchronisation, in turn, could lead to different dynamics such as normal and fractional diffusion.  Departing from a stochastic single cell trajectory, where a ``memory effect'' induces  persistent movement, we derive a kinetic-renewal system at the mesoscopic scale. We investigate various scenarios with different levels of complexity,  where the two ends of the cell move either independently or with partial or full synchronisation. We study the relevant macroscopic limits where we obtain  diffusion, drift-diffusion or fractional diffusion, depending on the initial system. This article clarifies the form of relevant macroscopic equations that describe the possible effects of synchronised movement in cells, and sheds light on the switching between normal and fractional diffusion.
\end{abstract}
\footnotetext[1]{B.P. has received funding from the European Research Council (ERC) under the European Union's Horizon 2020 research and innovation programme (grant agreement No 740623). G.E.R.  acknowledges the support of the Fondation Sciences Mathématiques de Paris (FSMP) for the postdoctoral fellowship.}

\section*{Introduction}\label{sec: intro}
Mathematical modelling of cell motility has been largely studied, specially through the use of partial differential equations (PDEs). Depending on the biological context, cell trajectories can be described by a persistent random walk \cite{othmer1988models}, where the individual tends to keep moving in the same direction as observed in  \cite{li2008persistent,fricke2016persistence}.
Often, directional persistence is also described as a correlated random walk where the direction of previous steps influences the direction of next ones. 
 
 This type of movement was studied in \cite{huda2018levy} where, due to the synchronisation of protrusions and retractions in the front and back of metastatic cells, the authors observed a strong presence of long runs, interspersed by a sequence of short steps.
 These characteristics are analogous to L\'{e}vy walk trajectories, where the probability of a long run, i.e. a trajectory in the same direction for a long time, is non negligible. It was also observed that in non-metastatic cells the front and back movements are independent and then they follow a classical random walk.
  In contrast to a Brownian motion, where the distribution of the individuals' trajectories follows a Gaussian, L\'{e}vy walk trajectories asymptotically follow a power-law distribution \cite{metzler2000random,zaburdaev2015levy}.
 Moreover, while for the Brownian case the mean square displacement $\langle x^2\rangle$ grows linearly with respect to time ($\langle x^2\rangle\sim t$), for the L\'{e}vy walk case we have $\langle x^2\rangle\sim t^{\zeta}$ where $\zeta\in(1,2)$. The exponent $\zeta=2$ corresponds to ballistic transport while $\zeta=1$ corresponds to normal diffusion. When $\zeta\in(1,2)$ we are in the superdiffusive regime. For the ubiquitous appearance of L\'{e}vy walk models in biological systems we refer to \cite{ariel2015swarming,harrisgeneralized,korobkova2004molecular} at the cellular level,  \cite{viswanathan1996levy,sims2008scaling,reynolds2017weierstrassian,focardi2009adaptive,reynolds2007displaced} for animals and  \cite{raichlen2014evidence,reynolds2018levy} for humans.

 In this work, we start from the most general description: when front and back can make independent, non-synchronised steps to the right and to the left. In this setting, the model records the persistence time in each direction, thus leading to a complex system which can better be understood in terms of cell elongation and movement of the center of gravity. 

To reduce the complexity, we assume that the cell length is fixed at the mesoscopic scale under investigation; the model is now amenable to multi-scale analysis and depending on the switching rates of both ends, we compute a diffusion coefficient at the macroscopic scale. Such a model with a given cell length can also be derived from the full system assuming fast switches in the dynamics of cell elongation compared to those for the movement.
We also investigate the synchronisation between back and front, where we can observe a full range of behaviours. With low persistence, normal diffusion occurs, possibly with a drift and explicit formulas are computed. They show the possibility of a ``backward diffusion'' regime which can be interpreted as instability of the constant steady state at the kinetic (mesoscopic) scale. More interesting is when persistence is higher, then fractional diffusion occurs exhibiting the phenomena of long excursions which motivates our study. 

When going back to individual cell (microscopic) dynamics, we perform numerical simulations and, as explained earlier, the long time behaviour quantifies the fractional diffusion in accordance to the developed theory.

To motivate our modelling, we are going to describe, based on biological evidence presented in \cite{huda2018levy}, two systems where none or full synchronisation of front and back leads to diffusive or superdiffusive dynamics, respectively.

\paragraph{Cell persistent movement through front-back synchronisation}\label{sec: movement description}

As previously introduced, the full synchronisation in space and time of the front and back movements of cells leads to L\'{e}vy walk dynamics, as observed in \cite{huda2018levy} for the case of metastatic cancer cells. On the other hand, when the cells' front and back movements  are independent, the trajectories follow a normal diffusion.

The {synchronisation} is translated into a persistence of the movement in a given direction. This means that at a given time step, the ends of the cell move forward or backwards simultaneously. The {non-synchronisation} is the result of independent movements in the front and back which might result in an intermittent cell movement pattern, where the cell length can vary. Hence, we assume that when the steps are not independent, and the cell keeps some ``{memory}'' of previous steps (non-Markovian process), the distribution of persistence length corresponds to a power-law and the movement is described by a L\'{e}vy walk. If the steps are independent, then the  trajectories  correspond to a diffusive movement. A more detailed description of this persistent and non-persistent movement is given in Section~\ref{sec: micro model} (see also Supplementary Information in \cite{huda2018levy}).

The aim of this article is to derive macroscopic equations that characterise these dynamics, starting from a kinetic description of the individual movement.
To study this behaviour we propose the following setting. We consider that a cell is approximated by a one dimensional system of two identical point masses  attached by an elastic cord. Each point mass is going to represent the front and back of the cell and the elastic cord represents the cell length. We aim to describe the trajectories of the whole system as in Figure \ref{fig: squetch}. Each mass is considered as a material point that takes discrete (in space and time) infinitesimally small steps to the left or to the right with a certain probability. For the non-synchronisation case (Figure~\ref{b}) we consider the independent movement of the front (represented by $y$) and back (represented by $x$) and the cell length change, where $\ell_c$ is the length at rest (see Subsection~\ref{subsec: biologically relevant switching prob}). While studying the synchronised movement leading to superdiffusion, since the front and back simultaneously move in the same direction at each time, we only consider the change in position of the centre of mass, represented by $x$ in Figure~\ref{a}. Since the cell length is fixed in this case, we do not take it into account.


\paragraph{Outline of the paper}

 In Section~\ref{sec: micro model} we present a detailed description of the individual movement as well as the main modelling assumptions. In Section~\ref{sec: de-synchronised movement} we introduce a general model where front and back movements are not synchronised. We also discuss some general notions as conservation of particles and realistic cell length for the model. Section~\ref{sec: simplified system with resting populations} presents a simplified version of the previous model where we fix the cell length at the mesoscopic scale. We derive macroscopic equations depending on the switching rates at the ends of the cell. This model is also obtained by considering a fast switching dynamics in the full system as described in Subsection~\ref{subsec: partial sync limit}. In Section~\ref{sec: syncrhonisation movement} we consider the system with synchronised front and back movement. We study the diffusive regime, when the persistence is low, and the superdiffusive regime in the opposite case (Section~\ref{sec: macroscopic equation for the synchronised movement}). Here we also show the possibility of a ``backwards diffusion'' regime. Finally, in Section~\ref{sec: numerics} we perform numerical simulations at the individual level to compare with the developed theory.

\section{Description of mathematical model}\label{sec: micro model}

\paragraph{Front-back non-synchronisation} In this case, the front and back movements are independent but with probabilities sampled from the same probability distribution. We consider that the front ($y$) gives a step $k_y$ of size $\delta$ at a given time $\tau$, and similarly for the back $x$. The cell length is given by $|y-x|$ where $\ell_c$ denotes the equilibrium length. 
If the whole system is moving to the right initially, the cell front is allowed to reverse direction and move to $y-\delta$ with probability $q_{k_y}=\tau p_{k_y}$ or it can keep moving in the same direction with probability  $\tilde{q}_{k_y}=1-\tau p_{k_y}$. Similarly, the back of the cell can move in the positive direction $x+\delta$ with probability $\tilde{q}_{k_x}=1-\tau p_{k_x}$ or reverse direction with probability $q_{k_x}=\tau p_{k_x}$. The length of the cell varies in a specific range so that we preserve the physical properties, as described later in Section \ref{subsec: biologically relevant switching prob}.

\begin{figure}[tbhp]
  \centering
\subfloat[\label{b}Non-synchronisation.]{\includegraphics[scale=0.6]{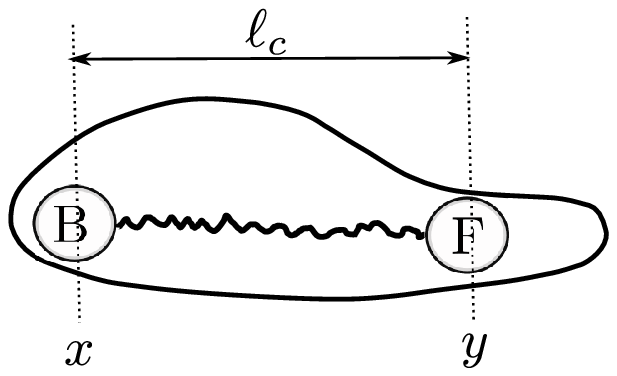}}
\hspace{1cm}
\subfloat[\label{a}Synchronisation]{\includegraphics[scale=0.6]{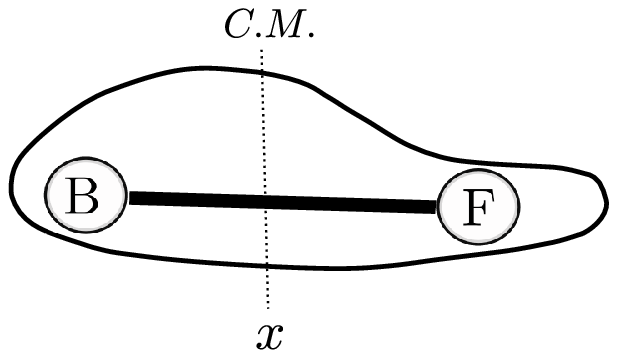}}
  \caption{Schematics of the front-back  cell movement.} \label{fig: squetch}
\end{figure}

\paragraph{Front-back synchronisation} For this case, since the front and back of the cell move simultaneously, i.e. at each time step they move to the left or to the right one step size with the same probability, the cell length is fixed at all times and we can consider the whole system as a point particle. The elastic cord can be considered as a solid road of fixed length and we study the movement of the center of mass only as in Figure \ref{a}. 

Analogous to the previous description, if we assume that the system is initially moving to the right, then it can reverse direction to $x-\delta$ with probability $q_k=\tau p_k$ or it can keep moving forward to $x+\delta$ with a probability $\tilde{q}_k=1-\tau p_k$. Every time the cell changes direction we set $k=0$ and we start counting again. The probability of keep moving without reversing will algebraically decrease with $k$. 

\noindent\emph{Note:} From now on, we use the notation $p_k$ when $k$ is discrete (see also Subsection~\ref{sec: sync movement}) and $p(k)$ when $k$ is continuous.

\paragraph{Reversing direction probability}
As described earlier, the rate at which the cell changes direction from left to right movement, given by $p_k$, depends on the number of steps $k$ given in that direction. The reversing rate is associated with a probability $\psi(k)$, which is given by
\begin{equation}
\psi(k)=e^{-\int_0^kp(k^*)\diff k^*}\ .\label{eq: survival probability}
\end{equation}
This function $\psi$ is often referred to as the \emph{survival probability}, i.e., it gives the probability that the event of interest, in this case the reverse in direction, has not occurred for $k$ steps. Equation~\eqref{eq: survival probability} means that the probability of moving for $k$ steps without changing is equal to the exponential of the cumulative reversing frequency.
As indicated in \cite{huda2018levy}, for the case of metastatic cells this probability decays algebraically with $k$, therefore, here we consider that 
\begin{equation}
    \psi(k)=\Bigl(\frac{k_0}{k_0+k}\Bigr)^\mu\ \ \ \textnormal{for}\ \ \mu\in (1,3)\ .\label{eq: psi}
\end{equation}
The reversing direction rate $p(k)$ can also be expressed as the ratio
\begin{equation}
    p(k)=\frac{\phi(k)}{\psi(k)}=\frac{-\partial_k \psi(k)}{\psi(k)}\ ,\label{eq: switching rate}
\end{equation}
where $\phi(k)$ is a probability density function. The above expression means that the reversing rate at step $k$ equals the density of the event divided by the probability of keep moving in the same direction for $k$ steps. 

\section{Non-synchronised movement description}\label{sec: de-synchronised movement}

\begin{figure}[ht!]
    \centering
    \includegraphics{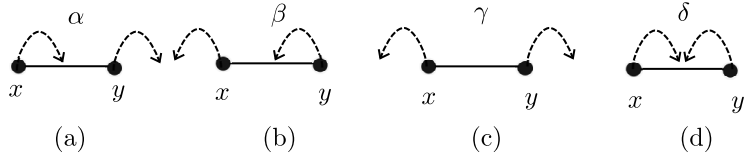}
    \caption{Non-synchronised movement where the cell moves to the right (a), to the left (b) or changes its length by stretching (c) and shrinking (d).}
    \label{fig: 2}
\end{figure}

We assume that the probabilities of the protrusions and retractions are independent from each other and therefore we have four different scenarios as in Figure~\ref{fig: 2}. We denote by $\alpha(t,x,y,k_x,k_y)$ and $\beta(t,x,y,k_x,k_y)$ the cells that are moving to the right and left, respectively, and by $\delta(t,x,y,k_x,k_y)$ and $\gamma(t,x,y,k_x,k_y)$ the cells that only change their length by elongation and contraction. Moreover, the steps at the cell front, denoted by $k_y$, are independent from the steps at the back, $k_x$, and consequently, we have to take into account the rates $p^\alpha_{x}$, $p^\alpha_{y}$, $p^\beta_{x}$, $p^\beta_{y}$, $p^\gamma_{x}$, $p^\gamma_{y}$, $p^\delta_{x}$ and $p^\delta_{y}$. These rates do not only depend on $k_x$ and $k_y$ but also on the distance $|x-y|$ to preserve the cell physical size, as we  discuss in  Section~\ref{subsec: biologically relevant switching prob}. When the front and back of the cell change direction simultaneously, $p_{x}=p_{y}$, we denote the corresponding switching rates as $p^\alpha,\ p^\beta,\ p^\gamma, \ p^\delta$. 

Note that cases (a) and (b) in Figure~\ref{fig: 2} are analogous to the synchronisation case later discussed in Section~\ref{sec: syncrhonisation movement}.

In full generality, the number density of each cell population is described by the following systems coupled through the boundary terms,
\begin{equation}
    \begin{aligned}
    \begin{cases}
    &(\partial_t+v\partial_x+v\partial_y+\partial_{k_x}+\partial_{k_y})\alpha=-(p_{x}^\alpha+p_y^\alpha)\alpha{-p^\alpha\alpha}\ ,\\
    &\alpha(t,x,y,k_x=0,k_y)=\int_0^\infty p_{x}^\gamma\gamma\diff k_x\ ,\\
    &\alpha(t,x,y,k_x,k_y=0)=\int_0^\infty p_{y}^\delta\delta\diff k_y\ ,\\
    & \alpha(t,x,y,k_x=0,k_y=0)=\int_0^\infty \int_0^\infty p^\beta\beta\diff k_x\diff k_y\ ,
    \end{cases}
    \end{aligned}
    \label{eq: right moving 1}
\end{equation}
\begin{equation}
    \begin{aligned}
     \begin{cases}
    &(\partial_t-v\partial_x-v\partial_y+\partial_{k_x}+\partial_{k_y})\beta=-(p_{x}^\beta+p^\beta_y)\beta{-p^\beta\beta}\ ,\\
    &\beta(t,x,y,k_x=0,k_y)=\int_0^\infty p_{x}^\delta\delta\diff k_x\ ,\\
    &\beta(t,x,y,k_x,k_y=0)=\int_0^\infty p_{y}^\gamma\gamma\diff k_y\ ,\\
    &\beta(t,x,y,k_x=0,k_y=0)=\int_0^\infty\int_0^\infty p^\alpha\alpha\diff k_x\diff k_y\ ,
    \end{cases}
    \end{aligned}
    \label{eq: left moving 1}
\end{equation}
\begin{equation}
    \begin{aligned}
     \begin{cases}
    &(\partial_t-v\partial_x+v\partial_y+\partial_{k_x}+\partial_{k_y})\gamma=-(p_{x}^\gamma+p_y^\gamma)\gamma{-p^\gamma\gamma}\ ,\\
    &\gamma(t,x,y,k_x=0,k_y)=\int_0^\infty p_{x}^\alpha\alpha\diff k_x\ ,\\
    &\gamma(t,x,y,k_x,k_y=0)=\int_0^\infty p_{y}^\beta\beta\diff k_y\ ,\\
   & \gamma(t,x,y,k_x=0,k_y=0)=\int_0^\infty\int_0^\infty p^\delta\delta\diff k_x\diff k_y\ ,
    \end{cases}
    \end{aligned}
    \label{eq: resting1}
\end{equation}
\begin{equation}
    \begin{aligned}
     \begin{cases}
    &(\partial_t+v\partial_x-v\partial_y+\partial_{k_x}+\partial_{k_y})\delta=-(p_{x}^\delta+p^\delta_y)\delta{-p^\delta\delta}\ ,\\
    &\delta(t,x,y,k_x=0,k_y)=\int_0^\infty p_{x}^\beta\beta\diff k_x\ ,\\
    &\delta(t,x,y,k_x,k_y=0)=\int_0^\infty p_{y}^\alpha\alpha\diff k_y\ ,\\
   &\delta(t,x,y,k_x=0,k_y=0)=\int_0^\infty\int_0^\infty p^\gamma\gamma\diff k_x\diff k_y\ .
    \end{cases}
    \end{aligned}
    \label{eq: resting2}
\end{equation}
The above systems of equations describe the different jumping combinations represented in Figure~\ref{fig: 2}. For instance, in the case of population $\alpha$, if $x$ changes direction with certain rate $p_x^\alpha$, then we have a transition from population $\alpha$ to $\gamma$ (Figure \ref{fig: 2} (a)$\to$(c)), which is given by $\gamma(t,x,y,k_x=0,k_y)$. Similarly, if $y$ changes direction in population $\alpha$ we have a change from $\alpha$ to $\delta$, and the individuals that leave population $\alpha$ appear in $\delta(t,x,y,k_x,k_y=0)$.  The reverse process also happens and transitions from population $\gamma$ to $\alpha$ (cells appear at $\alpha(t,x,y,k_x=0,k_y)$) and from $\delta$ to $\alpha$ (cells appear at $\alpha(t,x,y,k_x,k_y=0)$) are also considered. When the jumping direction changes at $x$ and $y$ simultaneously then it always happens that populations switch directly from $\alpha\longleftrightarrow \beta$ and $\gamma\longleftrightarrow\delta$.\\

\noindent\textbf{Notation:} For simplicity of notation, in the rest of the paper we use
\[
\iint\limits(\cdot)=\int_0^\infty\int_0^\infty(\cdot)\diff k_x\diff k_y\ .
\]
In the following we are going to check some physical properties that the systems \eqref{eq: right moving 1}-\eqref{eq: resting2} must satisfy.

\paragraph{Coordinates of the center of mass and cell elongation}

At a first stage, a desirable property is that cell polarization is preserved, that means $x<y$ all along the movement assuming it is true initially. To examine the conditions which enforce this property it is easier to use the coordinates of the center of mass and the distance between back and front. For that we let $\frac{x+y}{2}=X$ and $y-x=z$, the cell length. From now on, and for simplicity in the notation, we keep the same functions $\alpha, \ \beta,\ \gamma,\ \delta$ that will depend  on the new variables $(t,X,z,k_x,k_y)$.  From the system \eqref{eq: right moving 1}-\eqref{eq: resting2} we get
\begin{align}
    (\partial_t+2 v\partial_X+\partial_{k_x}+\partial_{k_y}){\alpha}&=-(p_{x}^\alpha+p^\alpha_y){\alpha}-p^\alpha{\alpha}\ , \label{eq: alpha cm coor}
    \\
   (\partial_t-2 v\partial_X+\partial_{k_x}+\partial_{k_y}){\beta}&=-(p_{x}^\beta+p_y^\beta){\beta}-p^\beta{\beta}\ ,\label{eq: beta cm coor}
   \\
    (\partial_t+ v\partial_z+\partial_{k_x}+\partial_{k_y}){\gamma}&=-(p_{x}^\gamma+p_y^\gamma){\gamma}-p^\gamma{\gamma}\ , \label{eq: gamma cm coor}
   \\    
    (\partial_t-\ v\partial_z+\partial_{k_x}+\partial_{k_y}){\delta}&=-(p_{x}^\delta+p_y^\delta) {\delta}-p^\delta{\delta}\ ,\label{eq: delta cm coor}
\end{align}
with the boundary conditions in $k_x$, $k_y$, 
\begin{align}
&{\alpha}(\cdot,k_x=0,k_y)= \int_0^\infty p_{x}^\gamma {\gamma} \diff k_x\ ,\quad
{\alpha}(\cdot,k_x,k_y=0)= \int_0^\infty p_{y}^\delta {\delta} \diff k_y\ , 
\quad {\alpha}(\cdot,k_x=0,k_y=0)=\iint\limits p^\beta {\beta}\ , 
\label{eq: alpha kx}
\\
&{\beta}(\cdot,k_x=0,k_y)= \int_0^\infty p_{x}^\delta {\delta} \diff k_x\ , \quad
{\beta}(\cdot,k_x,k_y=0)= \int_0^\infty p_{y}^\gamma {\gamma} \diff k_y\ , 
\quad {\beta}(\cdot,k_x=0,k_y=0)=\iint\limits p^\alpha {\alpha}\ ,
\label{eq: beta kx}
\\
&{\gamma}(\cdot,k_x=0,k_y)= \int_0^\infty p_{x}^\alpha {\alpha} \diff k_x\ , \quad
{\gamma}(\cdot,k_x,k_y=0)= \int_0^\infty p_{y}^\beta {\beta} \diff k_y\ ,  
\quad {\gamma}(\cdot,k_x=0,k_y=0)=\iint\limits p^\delta {\delta} \ ,
\label{eq: gamma kx}
\\
&{\delta}(\cdot,k_x=0,k_y)= \int_0^\infty p_{x}^\beta {\beta} \diff k_x\ ,\quad
{\delta}(\cdot,k_x,k_y=0)= \int_0^\infty p_{y}^\alpha {\alpha} \diff k_y\ ,
\quad {\delta}(\cdot,k_x=0,k_y=0)=\iint\limits p^\gamma{\gamma}\ .
\label{eq: delta kx}
\end{align}
Here $(\cdot)$ denotes the dependence on $(t, X, z)$. 
In order to guarantee that $z>0$ is preserved, we also need to ensure that ${\delta}(X,z=0,k_x,k_y)=0$ (similarly $\gamma(X, z=0,k_x,k_y)=0$), which means that the jump rate $p^\delta_x+ p^\delta_y \to \infty$ as $z\to 0$ and  
$\int_0^{\cdot}  (p^\delta_x+ p^\delta_y)(z) \diff z =\infty$. 
Using the notation in~\eqref{eq: specific transitions}, this means that 
$\int_0^{\cdot} \mu_{\delta}(z)\diff z =\infty$.

\paragraph{Conservation of particles} 
Integrating with respect to $k_x$ and $k_y$ we define the macroscopic density
\begin{equation}
\bar{{\alpha}}(t,X,z)=\iint\limits {\alpha}(\cdot,k_x,k_y)\ ,
\end{equation}
and similarly for $\bar{\beta},\ \bar{\gamma}$ and $\bar{\delta}$. Moreover, integrating with respect to $k_x$ and $k_y$ equations \eqref{eq: alpha cm coor}-\eqref{eq: delta cm coor} and adding them together we obtain the following macroscopic conservation equation
\begin{equation}
\partial_t(u+w)+2v\partial_X j+v\partial_zm=0\ .
\end{equation}
Here $u(t,X,z)=\bar{{\alpha}}+\bar{{\beta}}$ is the moving population, $w(t,X,z)=\bar{{\gamma}}+\bar{{\delta}}$ is the resting population, $j(t,X,z)=\bar{{\alpha}}-\bar{{\beta}}$ is the mean direction of motion and $m(t,X,z)=\bar{{\gamma}}-\bar{{\delta}}$ is the mean extension rate.

\subsection{Biologically relevant switching probabilities}\label{subsec: biologically relevant switching prob}

The switching rate is not going to depend only on the persisting steps $k_x$ and $k_y$  but also on the cell length $z$. Using  \eqref{eq: psi} and \eqref{eq: switching rate} we can write the general expression
\begin{equation}
p(k)=\frac{\mu}{1+k}\ .\label{eq: expression for p(k)}
\end{equation}
For the non-synchronised movement this rate is given by, for population $\alpha$ (and similarly for the rest),
\begin{equation}
p_x^\alpha(k_x)=\frac{\mu_\alpha(z)}{1+k_x}\ ,\qquad p_y^\alpha(k_y) = \frac{\mu_\alpha(z)}{1+k_y}\ .\label{eq: specific transitions}
\end{equation}
The dependence of $\mu$ on $z$ guarantees that we keep a realistic cell length as we will discuss below. For the synchronised case, since the cell does not change shape, we consider the switching rate given by~\eqref{eq: expression for p(k)}.

As described in \cite{huda2018levy} the length of a cell can only vary in a certain range. Considering that the resting length is $\ell_c=|y-x|$, we define $L_\textnormal{max}=3.5\ell_c$ and $L_\textnormal{min}=0.5\ell_c$ and the switching rates satisfy the following properties,
\begin{equation}
 p_{x}^\alpha=\begin{cases}
        \epsilon & \textnormal{if}\ z\gg L_\textnormal{max}\ ,\\ \frac{\mu_\alpha(z)}{1+k_x}& \textnormal{if} \ z\ll L_\textnormal{min}\ ,
    \end{cases}\ \ \ \
    p_{y}^\alpha=\begin{cases}
        \frac{\mu_\alpha(z)}{1+k_y}& \textnormal{if} \ z\gg L_\textnormal{max}\ ,\\ \epsilon & \textnormal{if}\ z\ll L_\textnormal{min}\ ,
    \end{cases}
    \label{eq: probability alpha}
\end{equation}
\begin{equation}
    p_{x}^\beta=\begin{cases}
         \frac{\mu_\beta(z)}{1+k_x} & \textnormal{if}\ z\gg L_\textnormal{max}\ ,\\ \epsilon & \textnormal{if} \ z\ll L_\textnormal{min}\ ,
    \end{cases}    \ \ \ \
    p_{y}^\beta=\begin{cases}
       \epsilon & \textnormal{if} \ z\gg L_\textnormal{max}\ ,\\  \frac{\mu_\beta(z)}{1+k_y} & \textnormal{if}\ z\ll L_\textnormal{min}\ ,
    \end{cases}
    \label{eq: probability beta}
\end{equation}
and finally,
\begin{equation}
    p^\gamma_{x}=\begin{cases}
       \frac{\mu_\gamma(z)}{1+k_x} & \textnormal{if} \ z\gg L_\textnormal{max}\ ,\\  \epsilon & \textnormal{if}\ z\ll L_\textnormal{min}\ ,
    \end{cases}\ \ \ \
    p^\gamma_{y}=\begin{cases}
         \frac{\mu_\gamma(z)}{1+k_y} & \textnormal{if}\ z\gg L_\textnormal{max}\ ,\\ \epsilon & \textnormal{if} \ z\ll L_\textnormal{min}\ ,
    \end{cases}
    \label{eq: probability resting}
\end{equation}
\begin{equation}
    p^\delta_{x}=\begin{cases}
       \epsilon & \textnormal{if} \ z\gg L_\textnormal{max}\ ,\\  \frac{\mu_\delta(z)}{1+k_x} & \textnormal{if}\ z\ll L_\textnormal{min}\ ,
    \end{cases}\ \ \ \
    p^\delta_{y}=\begin{cases}
         \epsilon & \textnormal{if}\ z\gg L_\textnormal{max}\ ,\\ \frac{\mu_\delta(z)}{1+k_y} & \textnormal{if} \ z\ll L_\textnormal{min}\ .
    \end{cases}
    \label{eq: probability resting2}
\end{equation}

Here $\epsilon$ is a small parameter. Let us take for instance the case when cells are moving right (population $\alpha$). In the limit when $z\gg L_\textnormal{max}$,  since we want to preserve the physical length of the cell, the front has to change direction with a higher rate ($p_y^\alpha$) while the back should keep moving in the same direction (small $p_x^\alpha$). On the other hand, in the limit when $z\ll L_\textnormal{min}$ the back of the cell has to switch direction ($p_x^\alpha$) while the front should keep moving without changing (small $p_y^\alpha$). 
The opposite happens when the cell is moving to the left (population $\beta$).

For the case when the cell is at rest ($\gamma$ and $\delta$), if $z\gg L_\textnormal{max}$ for the case of the population $\gamma$, the switching rate has to be very high at both ends so that the cell recovers the resting length $\ell_c$. If $z\ll L_\textnormal{min}$, then $p^\gamma_x$, $p_y^\gamma$ are very small. The opposite happens in the case of the population $\delta$.


\section{Simplified system with resting population}\label{sec: simplified system with resting populations}

{ The systems~\eqref{eq: alpha cm coor}-\eqref{eq: delta cm coor}  are very complex to analyse since they involve different dynamics such as left and right movement for four different populations, and additionally, the change in cell length.} Therefore, in this section we consider a simplified model of three populations: cells moving left ($\beta$), cells moving right ($\alpha$) and resting cells ($\gamma_0$).  The population $\gamma_0$ represents { the average of the populations $\gamma$ and $\delta$ described before assuming the mean cell length is constant, thus ignoring the variable~$z$.}
The dynamics are described now by
\begin{align}
&\begin{cases}
    &\e^2\partial_t\alpha_\e+\e v\partial_x\alpha_\e+\partial_k\alpha_\e=-{p}(k)\alpha_\e\ ,\\
    &\alpha_\e(t,0,x)=\zeta^\beta\int_0^\infty {p}(k)\beta_\e\diff k+\zeta^\gamma\gamma^\e_0(t,x,0)\ ,\label{eq: alpha in three populations}
    \end{cases}\\
    \vspace{0.5cm}
 &   \begin{cases}
    &\e^2 \partial_t\beta_\e- \e v\partial_x\beta_\e+\partial_k\beta_\e=-{p}(k)\beta_\e\ ,\\
    &\beta_\e(t,0,x)=\zeta^\alpha\int_0^\infty{p}(k)\alpha_\e\diff k+(1-\zeta^\gamma)\gamma^\e_0(t,x,0)\ , \label{eq: beta in three populations}
    \end{cases}\\
    \vspace{0.5cm}
  &  \begin{cases}
    &\e^2 \partial_t\gamma^\e_0-\partial_k\gamma_0^\e=(1-\zeta^\alpha){p}(k)\alpha_\e+(1-\zeta^\beta){p}(k)\beta_\e\ .\label{eq: gamma_0 scaled}
    \end{cases}
\end{align}
{The switching rate ${p}(k)$ describes both the change in direction of the center of mass and transition to rest state $\gamma_0$ where memory is  gradually lost. This replaces the movement of the front and the back of the cell as in Section~\ref{sec: de-synchronised movement} and is given by~\eqref{eq: expression for p(k)}.} Since we are not considering front and back movements then $\alpha$, $\beta$ and $\gamma_0$ depend only on $(t,x,k)$. In the above system, $\zeta\in(0,1)$ is a probability and note that we have introduced a diffusive scaling $(t,x)\mapsto(\bar{t}/\varepsilon^2,\bar{x}/\varepsilon)$. The individuals from population $\alpha$ that switch direction with rate ${p}(k)$ either start moving in the opposite direction with probability $\zeta^\alpha$, represented by the first term in $\beta_\e(t,x,0)$, or they go into a resting phase with probability $1-\zeta^\alpha$, given by the first term in the right hand side of~\eqref{eq: gamma_0 scaled}. A similar dynamic is followed by individuals in population $\beta$. On the other hand, individuals that are at rest, population $\gamma_0$,  start to move to the right, with probability $\zeta^\gamma$, or to the left with probability $1-\zeta^\gamma$.

As for the systems~\eqref{eq: right moving 1}-\eqref{eq: resting2}, we can easily check that~\eqref{eq: alpha in three populations}-\eqref{eq: gamma_0 scaled} preserves the number density of individuals.\\

When the rate ${p}(k)$ is large enough for large $k$, more precisely when $k e^{-\int_0^k {p}(k^*) \diff k^*}$ is integrable, then the large scale dynamics is normal diffusion. To explain this, we define the survival probability $\psi$ as
\begin{equation}
\psi(k) = Z e^{-\int_0^k {p}(k^*) \diff k^*}, \qquad Z^{-1} := \int_0^\infty e^{-\int_0^k {p}(k^*) \diff k^*} \diff k\ , \qquad \int_0^\infty {p}(k) \psi(k)\diff k = Z \ .\label{eq: definition Z}
\end{equation}

For ${p}(k)= \frac{\mu}{1+k}$ with $\mu >2$, then indeed $k e^{-\int_0^k {p}(k^*) \diff k^*}= \frac k{(1+k)^\mu}$ is integrable.
\\

 The first question is to determine under which conditions diffusion occurs in the small scale regime for $\varepsilon$. To do that, we compute the limiting $\alpha$, $\beta$, $\gamma_0$ as $\varepsilon\to 0$, using the solution of
\begin{align}
\begin{cases}
    &\partial_k \alpha=-{p}(k)\alpha\ ,\\
    &\alpha(t,x,0)=\zeta^\beta\int_0^\infty {p}(k) \beta\diff k+\zeta^\gamma\gamma_0(t,x,0)\ ,
  \\[10pt]
    &\partial_k \beta=-{p}(k)\beta\ ,\\
    &\beta(t,x,0)=\zeta^\alpha\int_0^\infty{p}(k)\alpha\diff k+(1-\zeta^\gamma)\gamma_0(t,x,0)\ ,\label{eq: resting case}
  \\[10pt]
    &-\partial_k \gamma_0=(1-\zeta^\alpha){p}(k)\alpha+(1-\zeta^\beta){p}(k)\beta\ .
    \end{cases}
\end{align}
Therefore, we obtain the limits
\begin{align}
    \alpha(t,x,k)&=\alpha(t,x,0)\frac{\psi(k)}{Z}=\bar{\alpha}(t,x)\psi(k)\ , \qquad
    \beta(t,x,k)=\beta(t,x,0)\frac{\psi(k)}{Z}=\bar{\beta}(t,x)\psi(k)\ , \label{eq: limit expressions alpha beta}
    \\
    \gamma_0(t,x,k)&= (1-\zeta^\alpha)\bar{\alpha}(t,x)\psi(k)+(1-\zeta^\beta)\bar{\beta}(t,x)\psi(k)\ , \nonumber
\end{align}
where $\bar{\alpha}(t,x)=\int_0^\infty\alpha\diff k$ and $\bar{\beta}(t,x)=\int_0^\infty\beta \diff k$.
 The $\gamma_0(t,x,0)$ is given by
 \begin{equation}
     \gamma_0(t,x,0)=(1-\zeta^\alpha)\bar{\alpha}(t,x)\psi(0)+(1-\zeta^\beta)\bar{\beta}(t,x)\psi(0)\ .\label{eq: initial condition gamma_0}
 \end{equation}
With these expressions, we can compute  a relation between   $\bar{\alpha}(t,x)$ and $\bar{\beta}(t,x)$ starting from 
\begin{align*}
    \bar{\alpha}(t,x)&=\frac{\alpha(t,x,0)}{Z}=\frac{\zeta^\beta}{Z}\int_0^\infty {p}(k)\beta(t,x,k)\diff k+\frac{\zeta^\gamma}{Z}\gamma_0(t,x,0)\\
    &=\zeta^\beta \bar{\beta}(t,x)+\zeta^\gamma[(1-\zeta^\alpha)\bar{\alpha}(t,x)+ (1-\zeta^\beta)\bar{\beta}(t,x)]\ ,
\end{align*}
that finally gives 
$
    \bar{\alpha}(t,x)=\frac{\zeta^\beta+\zeta^\gamma(1-\zeta^\beta)}{1-\zeta^\gamma(1-\zeta^\alpha)}\bar{\beta}(t,x).
$
Therefore, the condition for a diffusive limit, i.e., $\bar{\alpha}(t,x)=\bar{\beta}(t,x)$ turns out to be
\begin{equation}
\zeta^\gamma=\frac{1-\zeta^\beta}{2-\zeta^\alpha-\zeta^\beta}\ .\label{eq: zeta^gamma}
\end{equation}
Then, the second question is to compute the diffusion coefficient. The macroscopic conservation equation is obtained using
\[
\partial_t(\bar{\alpha}_\varepsilon+\bar{\beta}_\varepsilon+\bar{\gamma}_0^\varepsilon)+v\partial_xJ_\varepsilon=0\ ,\quad \textnormal{where}\quad J_\varepsilon\coloneqq\frac{\bar{\alpha}_\varepsilon-\bar{\beta_\varepsilon}}{\varepsilon}\ .
\]
The difficulty here is to compute the flux $J_\varepsilon$ when ${p}(k)$ is not constant. In the following we are going to compute the diffusive limit under the condition~\eqref{eq: zeta^gamma}, i.e., such that, as $\varepsilon\to 0$
\[
\bar{\alpha}(t,x)=\bar{\beta}(t,x)\ , \ \textnormal{for}\ \ \zeta^\gamma=\frac{1-\zeta^\beta}{2-\zeta^\alpha-\zeta^\beta}\ , \ \textnormal{and}\ \ J_\varepsilon\to -Dv\partial_x\bar{\beta}(t,x)\ .
\]
 
 We start by computing the Taylor expansion of $\alpha_\varepsilon$ using the equation
\[
\varepsilon^2\partial_t\frac{\alpha_\varepsilon(t,k,x)}{\psi(k)}+\varepsilon v\partial_x\frac{\alpha_\varepsilon(t,k,x)}{\psi(k)}+\partial_k\frac{\alpha_\varepsilon(t,k,x)}{\psi(k)}=0\ .
\]
Integrating with respect to $k$ we find, 
\begin{equation*}
\begin{aligned}
    \frac{\alpha_\e(t,x,k)}{\psi(k)}&=\frac{\alpha_\e(t,x,0)}{\psi(0)}-\e v\partial_x\int_0^k\frac{\alpha_\e(t,x,k^*)}{\psi(k^*)}\diff k^* +\mathcal{O}(\e^2)=
    \frac{\alpha_\e(t,x,0)}{\psi(0)}- \e k v\partial_x \frac{\alpha_\e(t,x,0)}{\psi(0)}+\mathcal{O}(\e^2)
    \\
    &=\frac{\alpha_\e(t,x,0)}{\psi(0)}-\e k v\partial_x \bar{\alpha}_\varepsilon +\mathcal{O}(\e^2)\ ,
\end{aligned}
\end{equation*}
since, integrating in $k$ after multiplying by $\psi(k)$, we find $\bar{\alpha}_\varepsilon= \frac{\alpha_\e(t,x,0)}{\psi(0)}+\mathcal{O}(\e)$. For $\beta_\varepsilon$, we obtain
\begin{equation}
\begin{aligned}
    \frac{\beta_\e(t,x,k)}{\psi(k)}&=\frac{\beta_\e(t,x,0)}{\psi(0)}+\e k v\partial_x \bar{\beta}_\varepsilon 
    +\mathcal{O}(\e^2)\ .\label{eq: taylor expansion alpha beta}
\end{aligned}
\end{equation}
From the above equations the aim is to compute
\begin{align}
    J_{\varepsilon}&= \frac{\bar{\alpha}_\varepsilon-\bar{\beta_\varepsilon}}{\varepsilon}
    =\int_0^\infty \psi(k)
    \frac{\alpha_\varepsilon(t,x,0)-\beta_\varepsilon(t,x,0)}{\varepsilon  \psi(0) }\diff k 
    -v\partial_x\int_0^\infty k \psi(k) (\bar{\alpha}_\varepsilon
    +\bar{\beta}_\varepsilon) \diff k +\mathcal{O}(\e) \nonumber
    \\ &
    = \frac{\alpha_\varepsilon(t,x,0)-\beta_\varepsilon(t,x,0)}{\varepsilon  \psi(0)}
    - v\partial_x (\bar{\alpha}_\varepsilon
    +\bar{\beta}_\varepsilon)\int_0^\infty k \psi(k)\diff k +\mathcal{O}(\e) \ .\label{eq: limiting flux}
\end{align}
In order to  compute the term $\alpha_\varepsilon(t,x,0)-\beta_\varepsilon(t,x,0)$, we first multiply by ${p}(k)\psi(k)$ in~\eqref{eq: taylor expansion alpha beta} and integrate in $k$. This gives, using that $\int_0^\infty k{p}(k)\psi(k) \diff k=1$,  
\begin{align}
    \int_0^\infty {p}(k)\beta_\e(t,x,k)\diff k&=\beta_\e(t,x,0)+\e v\partial_x \bar{\beta}_\varepsilon \int_0^\infty k{p}(k)\psi(k) \diff k  +\mathcal{O}(\e^2) \nonumber
    \\
    &=\beta_\e(t,x,0)+ \e v\partial_x \bar{\beta}_\varepsilon  +\mathcal{O}(\e^2)
    \ . \label{eq: auxiliary}
\end{align}
The boundary condition $\alpha_\e(t,x,0)$ in~\eqref{eq: alpha in three populations} can be written as, after using~\eqref{eq: initial condition gamma_0} and~\eqref{eq: auxiliary},
\begin{align*}
    \alpha_\e(t,x,0)&=(\zeta^\beta+\zeta^\gamma(1-\zeta^\beta))\int_0^\infty {p}(k)\beta_\e(t,x,k)\diff k+\zeta^\gamma(1-\zeta^\alpha)\int_0^\infty {p}(k)\alpha_\e(t,x,k)\diff k 
    \\
    &= (\zeta^\beta+\zeta^\gamma(1-\zeta^\beta)) [\beta_\e(t,x,0)+ \e v\partial_x \bar{\beta}_\varepsilon ]+ \zeta^\gamma(1-\zeta^\alpha)
    [\alpha_\e(t,x,0)- \e v\partial_x \bar{\alpha}_\varepsilon ] +\mathcal{O}(\e^2)\ ,
\end{align*}
and we obtain, for $c:=\frac{1}{1-\zeta^\gamma(1-\zeta^\alpha)}=\frac{1}{\zeta^\beta+\zeta^\gamma(1-\zeta^\beta)}$ thanks to~\eqref{eq: zeta^gamma},

\begin{align*}
\alpha_\e(t,x,0)&= \beta_\e(t,x,0)+\e v\partial_x \bar{\beta}_\varepsilon - c \zeta^\gamma(1-\zeta^\alpha)\e v\partial_x \bar{\alpha}_\varepsilon +\mathcal{O}(\e^2)
\\
&=\beta_\e(t,x,0)+\e v\partial_x \bar{\beta}_\varepsilon +(1-c) \e v\partial_x \bar{\alpha}_\varepsilon +\mathcal{O}(\e^2) \ .
\end{align*}
From here we obtain, with $\bar{\alpha}= \bar{\beta}$ the limits of $\bar{\alpha_\e}$ and $\bar{\beta_\e}$,
\begin{align*}   
\lim_{\e\to 0 }\frac{\alpha_\e(t,x,0)-\beta_\e(t,x,0)}{\e}&=(2-c) v\partial_x \bar{\alpha}\ .
\end{align*}
Finally, we write~\eqref{eq: limiting flux} and compute the diffusion coefficient
\begin{align}
    J&= 
      -v\partial_x\bar{\alpha}(t,x)\Bigl(2 \int_0^\infty\psi(k)k\diff k-\frac{2-c}{Z} \Bigr), \qquad D= 2 \int_0^\infty\psi(k)k\diff k-\frac{2-c}{Z} \ .
\end{align}

Note that, by opposition to the formalism developed in \cite{frankgoudon}, this diffusion coefficient is not always positive. This is because the authors also rescale $k$ in such a way to give more weights to the large values of $k$. When $D$ is negative, some instability arises for the kinetic model, which is analysed in Section~\ref{sec:stabilityAnalysis}.

\subsection{Partial synchronisation limit}\label{subsec: partial sync limit}

We may also assume that moving forward is more effective than elongating and shortening. To represent that, we may  derive a partial synchronisation limit starting from~\eqref{eq: alpha cm coor}-\eqref{eq: delta cm coor}. This limit consists on introducing fast transition rates for conformations $\gamma$ and $\delta$ so that the whole system approximately converges to the two moving populations $\alpha$ and $\beta$ (as in~\eqref{eq: alpha1} and~\eqref{eq: beta initial} in Section~\ref{sec: syncrhonisation movement}). We start by introducing the following scaling
\begin{align}
    (\partial_t+ v\partial_z+\partial_{k_x}+\partial_{k_y}){\gamma}&=-\frac{1}{\varepsilon} (p_{x}^\gamma+p_{y}^\gamma){\gamma}-p^\gamma{\gamma}\ ,
   \\    
    (\partial_t-\ v\partial_z+\partial_{k_x}+\partial_{k_y}){\delta}&=-\frac{1}{\varepsilon}(p_{x}^\delta +p_{y}^\delta){\delta}-p^\delta{\delta}\ ,
\end{align}
and we change accordingly the boundary conditions~\eqref{eq: alpha kx} and~\eqref{eq: beta kx}. With this scaling, we find ${\gamma}\to 0$ and ${\delta}\to 0$ as $\varepsilon \to 0$. The difficulty is to compute the limiting contribution to the boundary terms
\[
{\alpha}(\cdot,k_x=0,k_y)=\frac{1}{\varepsilon} \int_0^\infty p_{x}^\gamma {\gamma}(\cdot,k_x,k_y) \diff k_x\ .
\]
To do so, we use the method of characteristics in~\eqref{eq: gamma cm coor} and~\eqref{eq: delta cm coor} and neglect the initial contribution which is immediately absorbed due to our scaling. We find, respectively, 
\begin{equation}
{\gamma}(\cdot,k_x,k_y)=
\begin{cases}
    {\gamma}(t-k_x,X,z-vk_x,k_x=0,k_y-k_x)e^{-\int_0^{k_x}\Bigl(\frac{p_x^\gamma+p_y^\gamma}{\varepsilon} +p^\gamma\Bigr)\diff k_x^*} \quad \text{for}\quad  k_x<k_y\ ,
    \\
     {\gamma}(t-k_y,X,z-vk_y,k_x-k_y,k_y=0)e^{-\int_0^{k_y} \Bigl( \frac{p_x^\gamma+p^\gamma_y}{\varepsilon}+p^\gamma\Bigr)\diff k_y^*} \quad \text{for} \quad k_y <k_x\ ,
\end{cases}
\label{eq: charactersitic solution gamma}
\end{equation}
\begin{equation}
     {\delta}(\cdot,k_x,k_y)=\begin{cases}
    {\delta}(t-k_x,X,z+vk_x,k_x=0,k_y-k_x)e^{-\int_0^{k_x} \Bigl(\frac{p_x^\delta+p_y^\delta}{\varepsilon}+ p^\delta\Bigr)\diff k_x^*} \quad \text{for}\quad  k_x<k_y\ ,
    \\
     {\delta}(t-k_y,X,z+vk_y,k_x-k_y,k_y=0)e^{-\int_0^{k_y} \Bigl(\frac{p_x^\delta+p_y^\delta}{\varepsilon}+ p^\delta\Bigr)\diff k_y^*} \quad \text{for} \quad k_y <k_x\ .
\end{cases}
\label{eq: characteristic solution delta}
\end{equation}
{ We may estimate these quantities thanks to the Laplace approximation in the regime where $\varepsilon \to 0$,
\begin{equation}
\begin{aligned}
&\frac{1}{\varepsilon}\int_0^\infty p_x^\gamma \gamma(\cdot,k_x=0,k_y) e^{-\int_0^{k_x}\Bigl( \frac{p_x^\gamma+p_y^\gamma}{\varepsilon}+p^\gamma\Bigr)\diff k_x^*}\diff k_x \to \frac{p_x^\gamma(k_x=0)}{p_x^\gamma(k_x=0)+p_y^\gamma}\gamma(\cdot,k_x=0,k_y)\ ,\\
&\frac{1}{\varepsilon}\int_0^\infty p_x^\gamma\gamma(\cdot,k_x,k_y=0) e^{-\int_0^{k_y}\Bigl( \frac{p_x^\gamma+p_y^\gamma}{\varepsilon}+p^\gamma\Bigr)\diff k_y^*}\diff k_x\to {\delta_0(k_y)}\int_0^\infty \frac{p_x^\gamma}{p_x^\gamma+p_y^\gamma(k_y=0)}\gamma(\cdot,k_x,k_y=0)\diff k_x\ .\label{eq: exponential computation}
\end{aligned}
\end{equation}
}
{Note that $\delta_0(k_y)$ represents a Dirac delta function in $k_y$ while $\delta(\cdot,k_x,k_y)$ is the density of individuals moving as in Figure \ref{fig: 2}. Substituting these expressions into the first equation in~\eqref{eq: alpha kx} and using~\eqref{eq: gamma kx}, we find the boundary condition}
\begin{align}
{\alpha}(\cdot,k_x=0,k_y)
&\to  \frac{p_{x}^\gamma(k_x=0)}{p_{x}^\gamma(k_x=0)+p_{y}^\gamma}{\gamma}(\cdot,k_x=0,k_y)+
{\delta_0(k_y)}\int_0^\infty\frac{p_{x}^\gamma}{p_{x}^\gamma+p_{y}^\gamma (k_y=0)}{\gamma}(\cdot,k_x,k_y=0)\diff k_x\nonumber \\
&= \frac{p_{x}^\gamma(k_x=0)}{p_{x}^\gamma(k_x=0)+p_{y}^\gamma} \int_0^\infty p_x^\alpha  \alpha \diff k_x +{\delta_0(k_y)}\int_0^\infty\frac{p_{x}^\gamma}{p_{x}^\gamma+p_{y}^\gamma (k_y=0)} \int_0^\infty p_y^\beta  \beta \diff k_y \diff k_x\ .\label{eq: alpha kx=0}
\end{align}
Similarly, we obtain from the second equation in~\eqref{eq: alpha kx}
\begin{equation}
{\alpha}(\cdot,k_x,k_y=0)
=   {\delta_0(k_x)} \int_0^\infty \frac{p_{y}^\delta}{p_{x}^\delta(k_x=0)+p_{y}^\delta} \int_0^\infty p_x^\beta  \beta \diff k_x \diff k_y+ \frac{p_{y}^\delta(k_y=0)}{p_{x}^\delta+p_{y}^\delta (k_y=0)}\int_0^\infty p_y^\alpha  \alpha \diff k_y \ .\label{eq: alpha ky=0}
\end{equation}
We can simplify the above expressions by using the following notation
\begin{equation}
\begin{aligned}
\mathcal{P}_x^\gamma=\frac{p_x^\gamma}{p_x^\gamma+p_y^\gamma(k_y=0)}\ , \qquad \mathcal{P}_y^\gamma=\frac{p_x^\gamma(k_x=0)}{p_x^\gamma(k_x=0)+p_y^\gamma}\ ,
\\[15pt]
\mathcal{P}_x^\delta=\frac{p_y^\delta(k_y=0)}{p_y^\delta(k_y=0)+p_y^\delta}\ , \qquad \mathcal{P}_y^\delta=\frac{p_y^\delta}{p_x^\delta(k_x=0)+p_x^\delta}\ .
\label{eq: function P gamma delta}
\end{aligned}
\end{equation}

For the population ${\beta}$, we follow the same steps, starting from 
$
{\beta}(\cdot,k_x=0,k_y)=\frac{1}{\varepsilon}\int_0^\infty p_x^\delta{\delta}\diff k_x\ .
$
Using~\eqref{eq: exponential computation} we have
\begin{align}
    {\beta}(\cdot,k_x=0,k_y)&=\frac{p_x^\delta(k_x=0)}{p_x^\delta(k_x=0)+p_y^\delta}\int_0^\infty p_x^\beta{\beta}\diff k_x+\delta_0(k_y)\int_0^\infty\frac{p_x^\delta}{p_x^\delta+p_y^\delta(k_y=0)}\int_0^\infty p_y^\alpha
{\alpha}\diff k_y\diff k_x\ \label{eq: beta kx=0},\\
{\beta}(\cdot,k_x,k_y=0)&=\delta_0(k_x)\int_0^\infty\frac{p_y^\gamma}{p_x^\gamma(k_x=0)+p_y^\gamma}\int_0^\infty p_x^\alpha{\alpha}\diff k_x\diff k_y+\frac{p_y^\gamma(k_y=0)}{p_y^\gamma(k_y=0)+p_x^\gamma}\int_0^\infty p_y^\beta{\beta}\diff k_y\ .\label{eq: beta ky=0}
\end{align}
Here, we similarly define
\begin{equation}
\begin{aligned}
\bar{\mathcal{P}}_x^\gamma=\frac{p_y^\gamma(k_y=0)}{p_x^\gamma+p_y^\gamma(k_y=0)}\ , \qquad \bar{\mathcal{P}}_y^\gamma=\frac{p_y^\gamma}{p_x^\gamma(k_x=0)+p_y^\gamma}\ ,
\\[15pt]
\bar{\mathcal{P}}_x^\delta=\frac{p_x^\delta}{p_y^\delta(k_y=0)+p_y^\delta}\ , \qquad \bar{\mathcal{P}}_y^\delta=\frac{p_x^\delta(k_x=0)}{p_x^\delta(k_x=0)+p_x^\delta}\ .
\label{eq: function bar P gamma delta}
\end{aligned}
\end{equation}
Using~\eqref{eq: alpha kx=0},~\eqref{eq: alpha ky=0},~\eqref{eq: beta kx=0} and~\eqref{eq: beta ky=0} and integrating with respect to $k_x,\ k_y$ in~\eqref{eq: alpha cm coor} and~\eqref{eq: beta cm coor}
we obtain
\begin{align}
    (\partial_t&+2v\partial_X)\bar{{\alpha}} =-\iint\limits P_1(k_x,k_y)\alpha(\cdot,k_x,k_y)+\iint\limits P_2(k_x,k_y)\beta(\cdot,k_x,k_y)\ ,\label{eq: alpha limit}\\
    (\partial_t&-2v\partial_X)\bar{{\beta}} =-\iint\limits P_3(k_x,k_y)\beta(\cdot,k_x,k_y)+\iint\limits P_4(k_x,k_y)\alpha(\cdot,k_x,k_y) \ ,\label{eq: beta limit}
\end{align}
where $P_1,\ P_2,\ P_3,\ P_4$ are expressed in terms of~\eqref{eq: function P gamma delta} and~\eqref{eq: function bar P gamma delta}. This system is analogous to~\eqref{eq: alpha1}-\eqref{eq: beta initial} below that describes the left and right movement only, corresponding to the synchronisation case,  with a modified jumping rate $p(k)$ coming form the populations $\gamma$ and $\delta$.

To achieve conservation of particles we add~\eqref{eq: alpha limit} and~\eqref{eq: beta limit} to obtain
\begin{align}
    \partial_t u+2v\partial_X j=\iint\limits(-1+\mathcal{P}_y^\gamma +\bar{\mathcal{P}}_y^\gamma)p_x^\alpha{\alpha}&+\iint\limits(-1+\mathcal{P}^\delta_x +\bar{\mathcal{P}}_x^\delta )p_y^\alpha{\alpha}+ \iint\limits(-1+\mathcal{P}_x^\gamma+\bar{\mathcal{P}}_x^\gamma)p_y^\beta{\beta}\nonumber\\
    &+\iint\limits(-1+\mathcal{P}^\delta_y+\bar{\mathcal{P}}_y^\delta)p_x^\beta {\beta}=0\ ,
\end{align}
since the terms inside the brackets cancel out.

In conclusion, assuming fast transition in the ``asynchronous states'' $\gamma$ and $\delta$, we recover a simplified system where only the states $\alpha$, $\beta$ occur, that is the back and front are always synchronized. The new phenomena is the possibility of two fast transitions $\alpha \to \beta \to \alpha$ (or symmetrically exchanging $\beta $ and $\alpha$) which modifies the boundary conditions compared to the model initially postulated.

\section{Synchronised movement description}\label{sec: syncrhonisation movement}

 When the front and back movement of cells are synchronised, as described in Section~\ref{sec: micro model}, the system is much simpler but still can exhibit several remarkable features as oriented drift, instability or superdiffusive movement. As before,  we denote by $\alpha(t,x,k)$ the probability that the cell moves to the right and by $\beta(t,x,k)$  the probability that the cell moves to the left. 
The rate of changing the  direction is denoted by $p(k)$, defined in~\eqref{eq: expression for p(k)}. 
The system of equations that describes the synchronisation movement was derived in Appendix~\ref{sec: sync movement} from a discrete description and is given by

\begin{align}
&\begin{cases}
    &\partial_t\alpha(t,x,k)+v\partial_x\alpha(t,x,k)+\partial_k\alpha(t,x,k)=-p(k)\alpha(t,x,k)\ ,\\
    &\alpha(t,x,0)=\int_0^\infty p(k)\beta(t,x,k)\diff k\ ,\label{eq: alpha1}
    \end{cases}\\
    &\begin{cases}
    &\partial_t\beta(t,x,k)-v\partial_x\beta(t,x,k)+\partial_k\beta(t,x,k)=-p(k)\beta(t,x,k)\ ,\label{eq: beta initial}\\
&\beta(t,x,0)=\int_0^\infty p(k)\alpha(t,x,k)\diff k\ .
\end{cases}
\end{align}

\subsection{Normal diffusion limit of the synchronised system}\label{sec: diffusion}

We first study the scale in the memory term $p(k)$ which leads to a usual diffusion equation, following the lines of  Section~\ref{sec: simplified system with resting populations}. Because of the simplicity of the system, we may analyse the drift-diffusion behaviour in a more general context. To do so we re-scale~\eqref{eq: alpha1}-\eqref{eq: beta initial} as follows, 
\begin{equation}
\begin{cases}
  &\e^2  \partial_t\alpha_\e(t,x,k)+\e\, v\partial_x\alpha_\e(t,x,k)+\partial_k\alpha_\e(t,k,x)=-{(p(k)+ \e p^\alpha(k))} \alpha_\e(t,x,k)\ ,
  \\[5pt]
  &\e^2  \partial_t\beta_\e(t,x,k)-\e\, v\partial_x\beta_\e(t,x,k)+\partial_k\beta_\e(t,x,k)=-{(p(k)+ \e p^\beta(k))}\beta_\e(t,x,k)\ ,
  \\[5pt]
  &\alpha_\e(t,x,0)=\int_0^\infty {(p(k)+ \e p^\beta(k))}\beta_\e\diff k, \quad
  \beta_\e(t,x,0)=\int_0^\infty {(p(k)+ \e p^\alpha(k))}\alpha_\e\diff k\ .
\label{eq:alphaEps}
\end{cases}
\end{equation}
The drift-diffusion limit is obtained using again the identity 
\begin{equation}
\partial_t(\bar \alpha_\e+ \bar \beta_\e) +\partial_x J_\e =0\ , \qquad J_\e := \frac{\bar{\alpha_\e}-\bar{\beta_\e}}{\e} \ ,
\end{equation}
where we need to compute the $x$-flux $ J_\e$. We are going to compute the constants $V$ and $\tilde{D}$ such that, as $\e \to 0$,
\begin{equation}
\bar{\alpha}(t,x)= \bar \beta(t,x)\ , \quad \text{and} \quad J_\e \to V \bar \alpha(t,x) -\tilde{D}v \partial_x \bar \alpha(t,x)\ .
\label{eq_limitJ}\end{equation}
We complete the system~\eqref{eq:alphaEps} with initial data such that $\alpha_\e(0,x,k)=\beta_\e(0,x,k)$, this is because the definition of the flux $J_\e$ requires a bounded quantity $\frac{\bar{\alpha}_\e -\bar{\beta}_\e}{\e}$.

As $\e$ vanishes, we find limits that we denote by $\alpha, \; \beta$ and that satisfy 
\begin{equation*}
\begin{cases}
\partial_k\alpha(t,x,k)=-p(k)\alpha(t,x,k)\ ,\qquad  \alpha(t,x,0)= \int_0^\infty p(k)\beta(t,x,k)\diff k\ , 
  \\[5pt]
\partial_k\beta(t,x,k)=-p(k)\beta(t,x,k)\ ,\qquad  \beta(t,x,0)= \int_0^\infty p(k)\alpha(t,x,k)\diff k\ , 
\end{cases}
\end{equation*}
which means that the limits are given by~\eqref{eq: definition Z} and~\eqref{eq: limit expressions alpha beta}. 
 Consequently, we deduce that 
 \[\bar \alpha(t,x)=\frac{\alpha(t,x,0)}{\psi(0)}=\frac 1 {\psi(0)} \int_0^\infty p(k)\beta(t,x,k)\diff k=  \bar \beta(t,x) \ .
 \]
\\

To compute the flux $J_\varepsilon$ we follow the steps in Section~\ref{sec: simplified system with resting populations} and we write
\[
\frac{\alpha_\e(t,x,k)}{\psi(k)}=\frac{\alpha_\e(t,x,0)}{\psi(0)}- \e \int_0^k p^\alpha(k^*) \frac{\alpha_\e(t,x,k^*)}{\psi(k^*)}\diff k^*-
\e\, v\partial_x \int_0^k \frac{\alpha_\e(t,x,k^*)}{\psi(k^*)}\diff k^* +\mathcal{O}(\e^2)\ ,
\]
which can be re-arranged as
\begin{equation}
\alpha_\e(t,x,k)=\psi(k) \frac{\alpha_\e(t,x,0)}{\psi(0)}- \e \psi(k) \int_0^k p^\alpha(k^*) \diff k^* \frac{\alpha_\e(t,x,0)}{\psi(0)}-
\e\, k\psi(k) v\partial_x \frac{\alpha_\e(t,x,0)}{\psi(0)} +\mathcal{O}(\e^2)\ .
\label{eq_diff_alpha_eps}\end{equation}
Arguing in a similar way for $\beta_\e$, and using the expression for $\alpha_\varepsilon(t,x,0)$ in~\eqref{eq:alphaEps} we find
\begin{align}
\alpha_\e(t,x,0)=& \int_0^\infty {(p(k)+ \e p^\beta(k))}  \beta_\e \diff k= \beta_\e(t,x,0)\Bigl(1+\e \frac{Z^\beta}{\psi(0)}\Bigr) \nonumber
\\&\qquad - \e \frac{\beta_\e(t,x,0)}{\psi(0)} Y^{\beta}  
+ \e v\partial_x \frac{\beta_\e(t,x,0)}{\psi(0)} \ \int_0^\infty k p(k) \psi(k) \diff k + \mathcal{O}(\e^2)\ .\label{eq: alpha with Zbeta}
\end{align}
Here we have used the normalization of $\psi(k)$ introduced in~\eqref{eq: definition Z} and 
$$
Z^\alpha:= \int_0^\infty p^\alpha(k) \psi(k)\diff k ,\quad  \ Z^\beta:= \int_0^\infty p^\beta(k) \psi(k)\diff k, \quad   Y^{\alpha,\beta}:= \int_0^\infty p(k) \psi(k)\int_0^k p^{\alpha,\beta}(k^*) \diff k^* \diff k .
$$
We can further notice that $\int_0^\infty k p(k) \psi(k) \diff k=1$ as before. 

Writing an analogous expression of~\eqref{eq: alpha with Zbeta} but for $\beta_\varepsilon(t,x,0)$ we can compute
\begin{align*}
\alpha_\e(t,x,0)- & \beta_\e(t,x,0)  = \beta_\e(t,x,0) -\alpha_\e(t,x,0)+\e \frac{Z^\beta}{\psi(0)}\beta_\e(t,x,0)-\e \frac{Z^\alpha}{\psi(0)}\alpha_\e(t,x,0)
\\& -\e  \left[\frac{\beta_\e(t,x,0)}{\psi(0)} Y^{\beta}-\frac{\alpha_\e(t,x,0)}{\psi(0)} Y^{\alpha} \right]
 + \e v\partial_x   \left[\frac{\beta_\e(t,x,0)}{\psi(0)} + \frac{\alpha_\e(t,x,0)}{\psi(0)} \right] + \mathcal{O}(\e^2)\ .
\end{align*}
This shows that in the limit $\alpha(t,x,0)= \beta(t,x,0)$, $\bar{\alpha}(t,x)=\bar{\beta}(t,x)$ and thus $\alpha(t,x,k)= \beta(t,x,k)$, giving
\begin{align*}
2 \lim_{\e \to 0} \frac{\alpha_\e(t,x,0)- \beta_\e(t,x,0)}{\e}&= (Z^\beta-Z^\alpha) \bar{\alpha}(t,x) -\bar{\alpha}(t,x)(Y^\beta-Y^\alpha) +2 v\partial_x\bar{\alpha}(t,x)
\\&= 2 v\partial_x\bar{\alpha}(t,x) \ ,
\end{align*}
since (and the same argument shows that $Y^\beta=Z^\beta$),
\begin{align*}
Y^\alpha &= 
- \int_0^\infty \frac{d\psi(k)}{dk}  \int_0^k p^\alpha(k^*)\diff k^*=\int_0^\infty \psi(k) p^\alpha(k)\diff k= Z^\alpha.
\end{align*}

Back to~\eqref{eq_diff_alpha_eps}, using~\eqref{eq: limit expressions alpha beta}, we find, with $V= \int_0^\infty \psi(k) \left(\int_0^k (p^\alpha(k^*)-p^\beta(k^*))\diff k^*\right)\diff k$,
\begin{align*}
\lim_{\e \to 0} J_\e  & = \lim_{\e \to 0} \frac{\alpha_\e(t,x,0)- \beta_\e(t,x,0)}{\e \psi(0)}- \bar \alpha(t,x) \int_0^\infty \psi(k) \left(\int_0^k (p^\alpha(k^*)-p^\beta(k^*))\diff k^*\right)\diff k
  \\&\quad - 2 v\partial_x \bar \alpha (t,x)\int_0^\infty k \psi(k) \diff k
 = V \bar \alpha(t,x) - 2 v\partial_x\bar \alpha (t,x) \int_0^\infty k \psi(k)\left(1 -\frac{p(k)}{2Z}\right) \diff k\ . 
\end{align*}
We finally obtain the transport coefficients $V$ and $\widetilde D$
\[
\lim_{\e \to 0} J_\e = V \bar \alpha -  v \widetilde D \partial_x\bar \alpha (t,x), \qquad \widetilde D= 2\int_0^\infty k \psi(k)\diff k-\frac {1}{Z} \ .
\]

We recover the result of Section~\ref{sec: simplified system with resting populations} when $p^{\alpha, \beta}=0$, and $\zeta^{\alpha, \beta}=1$, then we find $V=0$, $c=1$ and $D=\widetilde D$. The same comment on the positivity of $\widetilde D$ applies here.\\

\begin{rem}
When $\psi(k)$ is given by~\eqref{eq: psi}, then for $k_0=1$ we have that $\int_0^\infty k\psi(k)\diff k=\frac{1}{(\mu-2)(\mu-1)}>0$ for $\mu>2$. Moreover, it holds that
\[
\frac{1}{(\mu-2)(\mu-1)}>\frac{1}{2Z}\qquad\textnormal{for}\qquad \mu\in(2,3)\ 
\]
 and the diffusion coefficient $\tilde{D}>0$.
\end{rem}

\subsection{Stability analysis}
\label{sec:stabilityAnalysis}

Consider, for simplicity, the system~\eqref{eq:alphaEps}. When the above diffusion coefficient is negative, $\widetilde D<0$, we expect instability for the kinetic system when $\e$ small enough. This phenomena has been already observed for chemotaxis and semilinear parabolic equations in \cite{Pyasuda2018,Pmoussa2019} and we explain it in the present context.

To study the stable/unstable modes, we consider a simple Fourier mode 
$
\alpha(t,x,k)=\bar{\alpha}(t,x)+a_n(k)e^{\lambda t}e^{inx}$
that we substitute in the equation for $\alpha$ in~\eqref{eq:alphaEps}. We get
\[
\e^2\lambda a_n(k)+\e vina_n(k)+\partial_ka_n(k)+p(k)a_n(k)=0\ ,
\]
which we can solve to obtain
\[ \begin{cases}
a_n(k)=a_n(0)e^{-\int_0^k(p(k^*)+\e^2\lambda+\e ivn)\diff k^*}\ ,
\\[5pt]
a_n(0)=\int_0^\infty p(k)b_n(k)\diff k\ .
\end{cases}\]
Hence we have
\begin{align*}
    a_n(k)& = \int_0^\infty p(k)b_n(k)\diff k\ e^{-\int_0^k(p(k^*)+\e^2\lambda+\e ivn)\diff k^*}\ ,\\
    b_n(k) &= \int_0^\infty p(k)a_n(k)\diff k\ e^{-\int_0^k(p(k^*)+\e^2\lambda-\e ivn)\diff k^*}\ .
\end{align*}
Substituting $b_n(k)$ into $a_n(k)$, multiplying by $p(k)$ and integrating we obtain the {\em dispersion relation}
\begin{equation}
    1=\int_0^\infty p(k)e^{-\int_0^k(p(k^*)+\e^2\lambda+\e ivn)\diff k^*}\diff k\int_0^\infty p(k)e^{-\int_0^k(p(k^*)+\e^2\lambda-\e ivn)\diff k^*}\diff k\ .\label{eq: before Taylor}
\end{equation}
For $\e$ very small, we use a Taylor expansion and re-write~\eqref{eq: before Taylor} as
\begin{align*}
    1 & =\int_0^\infty p(k)e^{-\int_0^k p(k^*)\diff k^*}(1-\e ivnk-\e^2\lambda k-\frac{\e^2}{2}n^2k^2v^2)\diff k\nonumber\\ &\ \ \ \times\int_0^\infty p(k)e^{-\int_0^kp(k^*)\diff k^*}(1+\e ivnk-\e^2\lambda k-\frac{\e^2}{2}n^2k^2v^2)\diff k\ .
\end{align*}
As before, see~\eqref{eq: definition Z}, we may use that $\int_0^\infty p(k)e^{-\int_0^k p(k^*)\diff k^*}\diff k=1$, $\int_0^\infty k p(k)e^{-\int_0^k p(k^*)\diff k^*}\diff k=\frac 1 Z$. Thus
the terms of order $1$ and $\e$ cancel and the second order terms give
\begin{align*}
\frac 2 Z \lambda = -n^2v^2\int_0^\infty p(k)e^{-\int_0^kp(k^*)\diff k^*}k^2\diff k +\frac 1{Z^2} v^2n^2\ .
\end{align*}
Computing 
$\int_0^\infty p(k)\psi(k)k^2\diff k=2\int_0^\infty\psi(k)k\diff k$, 
we obtain
\begin{equation}
    \lambda = \frac{n^2v^2}{2}\left(\frac{1}{Z}-2\int_0^\infty\psi(k)k\diff k \right)= - \widetilde D\ .
\end{equation}
This condition shows that when  $\widetilde D<0$, for $\e$ small, the kinetic model is Turing unstable as in \cite{Pyasuda2018,Pmoussa2019}.

\section{Fractional equation for the synchronised movement} 
\label{sec: macroscopic equation for the synchronised movement}

In the context of system~\eqref{eq: alpha1}-\eqref{eq: beta initial}, using an appropriate scaling, we obtain a macroscopic fractional diffusion equation describing the persistent movement of the total population when the front and back of the cell are synchronised. When $\psi(k)$ has a ``fat tail'', meaning $\mu\in(1,2)$ for $p(k)= \frac \mu {1+k}$, fractional diffusion occurs as already pointed out in~\cite{franksun2018}. We also recall that superdiffusion regimes are well established in different contexts of purely kinetic theory since the seminal works \cite{Olla2009,M3}.

\subsection{Kinetic system} 
We start by integrating~\eqref{eq: alpha1} and~\eqref{eq: beta initial} with respect to $k$, taking into account the boundary condition at $k=0$,
\begin{align}
\partial_t\bar{\alpha}(t,x)+v\partial_x\bar{\alpha}(t,x)&=-\int_0^{t} p(k)\alpha(t,x,k)\diff k +\int_0^{t} p(k)\beta(t,x,k)\diff k\ ,\label{eq: alpha bar}\\
\partial_t\bar{\beta}(t,x)-v\partial_x\bar{\beta}(t,x)&=-\int_0^{t} p(k)\beta(t,x,k)\diff k+\int_0^{t} p(k) \alpha(t,x,k)\diff k\ , \label{eq: beta bar}
\end{align}
where 
$\bar{\alpha}(t,x)=\int_0^{t}\alpha(t,x,k)\diff k$ {and} $\bar{\beta}(t,x)=\int_0^{t}\beta(t,x,k)\diff k\ .
$ We also consider initial conditions
$\beta^0(0,x,k)=\bar{\beta}^0(x)\delta(k)
$ and $\alpha^0(0,x,k)=\bar{\alpha}^0(x)\delta(k)$.\\

Now the aim is to write the right hand side of~\eqref{eq: alpha bar}-\eqref{eq: beta bar} in terms of the macroscopic densities $\bar{\alpha}(t,x)$ and $\bar{\beta}(t,x)$. For that purpose we follow some steps from \cite{estrada2018fractional} and  \cite{fedotov2015persistent}.
Using the method of characteristics we find the solution of~\eqref{eq: alpha1} and~\eqref{eq: beta initial} for $k<t$ {where we neglect the initial data}:
\begin{align}
    \alpha(t,x,k)&=\alpha(t-k, x-vk,0)e^{-\int_0^kp(k^*)\diff k^*}\ \label{eq: characteristic alpha},\\
    \beta(t,x,k)&=\beta(t-k,x+vk,0)e^{-\int_0^k p(k^*)\diff k^*}\ .\label{eq: characteristic beta}
\end{align}
Next, from~\eqref{eq: alpha bar} let us define the escape and arrival rates of individuals at position $x$ at time $t$ as 
\begin{equation}
j_\alpha(t,x)=\int_0^{t}p(k)\alpha(t,x,k)\diff k\ ,\qquad j_\beta(t,x)=\int_0^{t}p(k)\beta(t,x,k)\diff k\ .\label{eq: switching rates j}
\end{equation}
Recalling the definitions~\eqref{eq: survival probability} and~\eqref{eq: switching rate} and following the steps in Appendix~\ref{sec: escaping and arrival rates}-I we write
\begin{align}
    j_\alpha(t,x)&=\int_0^{t}\phi(t-s)e^{-v(t-s)\partial_x}\alpha(s,x,0)\diff s+{\alpha}^0(x-vk)\phi(k)\ ,\label{eq: def j_alpha}    \\
j_\beta(t,x)&=\int_0^{t}\phi(t-s)e^{v(t-s)\partial_x}\beta(s,x,0)\diff s+{\beta}^0(x+vk)\phi(k)\ .\label{eq: def j_beta}
\end{align}
Using the Laplace transform $\mathcal{L}[f](t)=\hat{f}(\lambda)$ where $\lambda$ is the Laplace variable, we have
\begin{equation}
    \hat{j}_\alpha(\lambda,x)=\hat{\phi}(\lambda+v\partial_x)\hat{\alpha}(\lambda,x,0)\, +{\alpha}^0\hat{\phi}(\lambda+v\partial_x)\ .\label{eq: j laplace}
\end{equation}
Moreover, using the Laplace transform of the characteristic solution~\eqref{eq: characteristic alpha}  and the definition of $\bar{\alpha}(t,x)$ we write
\begin{equation}
    \hat{\bar{\alpha}}(\lambda,x)=\hat{\alpha}(\lambda,x,0)\hat{\psi}(\lambda+v\partial_x)+{\alpha}^0\hat{\psi}(\lambda+v\partial_x)\ .\label{eq: beta laplace}
\end{equation}
Substituting $\hat{\alpha}(\lambda,x,0)$ from~\eqref{eq: beta laplace} into~\eqref{eq: j laplace} we finally get
\begin{equation}
\hat{j}_\alpha(\lambda,x)=\frac{\hat{\phi}(\lambda+v\partial_x)}{\hat{\psi}(\lambda+v\partial_x)}\hat{\bar{\alpha}}(\lambda,x)\, =\hat{Q}(\lambda+v\partial_x)\hat{\bar{\alpha}}(\lambda,x)\, \ .\label{eq: memory kernel definition}
\end{equation}
The operator $\hat{Q}(\lambda+v\partial_x)$  can be explicitly computed in the Laplace space.
Transforming back to the $(t,x)$-space we have, for $j_\alpha$ and $j_\beta$,
\begin{align}
      j_\alpha(t,x)&=\int_0^{t}{Q}(t-s)\bar{\alpha}(s,x-v(t-s))\diff s\label{eq: j alpha}\ ,\\
  j_\beta(t,x)&=\int_0^{t} Q(t-s)\bar{\beta}(s,x+v(t-s))\diff s\label{eq: j beta}\ .
\end{align}
Using the expressions~\eqref{eq: j alpha} and~\eqref{eq: j beta}  we write the system~\eqref{eq: alpha bar}-\eqref{eq: beta bar} in term of the macroscopic quantities $\bar{\alpha}$ and $\bar{\beta}$. 
In the following we obtain explicit expressions for $j_\beta$ and $j_\alpha$ by using the distribution of persistence steps $k$ given in~\eqref{eq: psi}.

\subsection{Left and right persistent movement}\label{sec: left and right persistence}
Using the results from the previous section we write the kinetic system as follows
\begin{align}
    \partial_t\bar{\alpha}+v\partial_x\bar{\alpha}&=-j_\alpha+j_\beta\label{eq: kinetic 1}\ ,\\
    \partial_t\bar{\beta}-v\partial_x\bar{\beta}&=-j_\beta+j_\alpha\ ,\label{eq: kinetic 2}
\end{align}
where $j_\alpha$ and $j_\beta$ are given by~\eqref{eq: j alpha} and~\eqref{eq: j beta}, respectively.

The quantities in the right hand side of~\eqref{eq: kinetic 1} and~\eqref{eq: kinetic 2} are best expressed in the Fourier-Laplace space, where the Fourier-Laplace transform is defined as 
\begin{equation*}
\mathcal{F L}[f](t,x)=\tilde{f}(\lambda,\xi)=\int_{\mathds{R}}\int_0^\infty e^{i\xi x-\lambda t}f(t,x)\diff t\diff x \ .
\end{equation*}
 Transforming the system~\eqref{eq: kinetic 1}-\eqref{eq: kinetic 2}
we write
\begin{equation}\begin{aligned}
    \lambda\tilde{\bar{\alpha}}+\alpha^0+ vi\xi\tilde{\bar{\alpha}}=-\tilde{j}_\alpha+\tilde{j}_\beta\ ,\\
    \lambda\tilde{\bar{\beta}}+\beta^0- vi\xi\tilde{\bar{\beta}}=-\tilde{j}_\beta+\tilde{j}_\alpha\ ,\label{eq: scaled system}
\end{aligned}\end{equation}
where
$\tilde{j}_\alpha=\tilde{Q}(\lambda+ vi\xi )\tilde{\bar{\alpha}}(\lambda,\xi)$ and $ \tilde{j}_\beta=\tilde{Q}(\lambda- vi\xi )\tilde{\bar{\beta}}(\lambda,\xi)$.
To obtain $\tilde{Q}(\lambda\pm vi\xi)$ we first have to compute the quantities $\tilde{\phi}(\lambda\pm vi\xi)$ and $\tilde{\psi}(\lambda\pm vi\xi)$, previously defined in~\eqref{eq: psi} and~\eqref{eq: switching rate}.  Letting $\lambda_\pm=\lambda\pm vi\xi$, $\tilde{\phi}^\pm=\tilde{\phi}(\lambda_\pm)$ and $\tilde{\psi}^\pm=\tilde{\psi}(\lambda_\pm)$ we write, 
\begin{equation*}
    \tilde{\psi}^\pm=k_0^\mu\lambda_\pm^{\mu+1}e^{k_0\lambda_\pm}\Gamma(-\mu+1,k_0\lambda_\pm)\ ,\qquad \tilde{\phi}^\pm=\mu(k_0\lambda_\pm)^\mu\Gamma(-\mu,k_0\lambda_\pm)e^{k_0\lambda_\pm}\ .
\end{equation*}
Using an asymptotic expansion of the Gamma function \cite{NIST:DLMF} and following the steps in \cite{estrada2018fractional} we get
\begin{equation}
    \begin{aligned}
    \tilde{\psi}^\pm&=-\frac{k_0}{1-\mu}-\frac{k_0^2\lambda_\pm}{(1-\mu)(2-\mu)}+k_0^\mu\lambda_\pm^{\mu-1}\Gamma(-\mu+1)+\mathcal{O}(k_0^3\lambda_\pm^2)\ ,\\
    \tilde{\phi}^\pm &=1+\frac{k_0\lambda_\pm}{1-\mu}+k_0^\mu\lambda_\pm^{\mu}+\mathcal{O}(k_0^{\mu+1}\lambda_\pm^{\mu+1})\ .\label{eq: important expansions}
\end{aligned}
\end{equation}
Using~\eqref{eq: important expansions} we can write
\begin{equation}
    \tilde{Q}(\lambda\pm vi\xi)\simeq\frac{\mu-1}{k_0}-\frac{\lambda\pm vi\xi}{2-\mu}-k_0^{\mu-2}(\lambda\pm vi\xi)^{\mu-1}(\mu-1)\Gamma(-\mu+1)\ .
\end{equation}
Hence, system~\eqref{eq: scaled system} is now written in the $(t,x)$-space, for $b=k_0^{\mu-1}(\mu-1)\Gamma(-\mu+1)$,
\begin{align*}
    \partial_t\bar{\alpha}+ v\partial_x\bar{\alpha}&=-\frac{\mu-1}{k_0}(\bar{\alpha}-\bar{\beta})+\frac{\partial_t+v\partial_x}{2-\mu}\bar{\alpha}-\frac{\partial_t-v\partial_x}{2-\mu}\bar{\beta} +b\Bigl( (\partial_t+v\partial_x)^{\mu-1}\bar{\alpha}-(\partial_t-v\partial_x)^{\mu-1}\bar{\beta}\Bigr)\ ,\\
    \partial_t\bar{\beta}- v\partial_x\bar{\beta}&=-\frac{\mu-1}{k_0}(\bar{\beta}-\bar{\alpha})+\frac{\partial_t-v\partial_x}{2-\mu}\bar{\beta}-\frac{\partial_t+v\partial_x}{2-\mu}\bar{\alpha} +b\Bigl((\partial_t-v\partial_x)^{\mu-1}\bar{\beta}-(\partial_t+v\partial_x)^{\mu-1}\bar{\alpha}\Bigr)\ .
\end{align*}

Here we have used the fact that
\begin{equation*}
    \mathcal{FL}\Bigl[ \Bigl(\partial_t\pm v\partial_x\Bigr)^{\mu-1} f\Bigr]=(\lambda\pm vi\xi)^{\mu-1}\tilde{f}\ .
\end{equation*}
\begin{rem}
The tempered fractional material derivative   \cite{sokolov2003towards,fedotov2015persistent}, defined as
\[
\Bigl(\partial_t\pm v\partial_x \Bigr)^{\mu-1}f(t,x)={}_{0}D^{\mu-1}_tf(t,x\pm vt)\ ,
\]
generalises the standard material derivative $\frac{\diff }{\diff t}f(t,x\pm vt)=(\partial_t\pm v\partial_x)f$ for $\mu=2$  through the introduction of the Riemann-Liouville operator \cite{sokolov2003towards}.
\end{rem}

\subsection{Macroscopic PDE for the total population}

We may now write a macroscopic equation for the total density $\rho(t,x)=\bar{\alpha}(t,x)+\bar{\beta}(t,x)$.
From the definitions of $j_\alpha$ and $j_\beta$ in~\eqref{eq: switching rates j} we know that
\begin{align}
    j_\alpha(t,x)=\beta(t,x,0)\ \ \textnormal{and}\ \ j_\beta(t,x)=\alpha(t,x,0)\ ,
\end{align}
and therefore, using~\eqref{eq: characteristic alpha} and~\eqref{eq: characteristic beta} we have
\begin{equation}
    \alpha(t,x,k)=j_\beta(t-k,x-vk)\psi(k)\ ,\ \ \ \beta(t,x,k)=j_\alpha(t-k,x+vk)\psi(k)\ .
    \label{eq: other definition}
\end{equation}
Hence, from~\eqref{eq: other definition} we can write
\begin{align}
    j_\alpha(t,x)&=\int_0^{t}p(k)\alpha(t,x,k)\diff k=\int_0^{t}\phi(k)j_\beta(t-k,x-vk)\diff k+{\alpha}^0(x-vk)\phi(k)\label{eq: j_alpha 1}\ ,\\ j_\beta(t,x)&=\int_0^{t}p(k)\beta(t,x,k)\diff k=\int_0^{t}\phi(k)j_\alpha(t-k,x+vk)\diff k+{\beta}^0(x+vk)\phi(k) \label{eq: j_beta 1}\ .
\end{align}
On the other hand we have
\begin{align}
    \bar{\alpha}(t,x)&=\int_0^{t}\alpha(t,x,k)\diff k=\int_0^{t}j_\beta(t-k,x-vk)\psi(k)\diff k+{\alpha}^0(x-vk)\psi(k)\ ,\label{eq: alpha 1}\\
    \bar{\beta}(t,x)&=\int_0^{t}\beta(t,x,k)\diff k=\int_0^{t}j_\alpha(t-k,x+vk)\psi(k)\diff k+{\beta}^0(x+vk)\psi(k)\ .\label{eq: beta 1}
\end{align}

Next we apply the Fourier-Laplace transform to~\eqref{eq: j_alpha 1}-\eqref{eq: beta 1}  to obtain
\begin{align}
    \tilde{j}_\alpha(\lambda,\xi)&=\tilde{\phi}(\lambda+iv\xi)\Bigl(\tilde{j}_\beta(\lambda,\xi)+\tilde{\alpha}^0(\xi) \Bigr)\ ,\ \  \tilde{j}_\beta(\lambda,\xi)=\tilde{\phi}(\lambda-iv\xi)\Bigl(\tilde{j}_\alpha(\lambda,\xi)+\tilde{\beta}^0(\xi)\Bigr)\ ,\\
    \vspace{0.1cm}
    \tilde{\bar{\alpha}}(\lambda,\xi)&=\Bigl( \tilde{j}_\beta(\lambda,\xi)+\tilde{\alpha}^0(\xi)\Bigr)\tilde{\psi}(\lambda+iv\xi)\ ,\ \ 
    \tilde{\bar{\beta}}(\lambda,\xi)=\Bigl( \tilde{j}_\alpha(\lambda,\xi)+\tilde{\beta}^0(\xi)\Bigr)\tilde{\psi}(\lambda-iv\xi)\ .\label{eq: beta tilde bar}
\end{align}

 Re-arranging the above expressions and using the notation introduced in Section \ref{sec: left and right persistence} for $\psi^{\pm}$ and $\phi^\pm$ we get
\begin{equation}
    \tilde{\bar{\alpha}}(\lambda,\xi)=\Bigl(\frac{\tilde{\phi}^-}{\tilde{\psi}^-}\tilde{\bar{\beta}}(\lambda,\xi)+\tilde{\alpha}^0 \Bigr)\tilde{\psi}^+\ ,\ \ \tilde{\bar{\beta}}(\lambda,\xi)=\Bigl(\frac{\tilde{\phi}^+}{\tilde{\psi}^+}\tilde{\bar{\alpha}}(\lambda,\xi)+\tilde{\beta}^0 \Bigr)\tilde{\psi}^-\ ,\label{eq: alpha and beta important}
\end{equation}
or equivalently,
\begin{equation}
    \tilde{\bar{\alpha}}(\lambda,\xi)=\frac{\tilde{\phi}^-\tilde{\beta}^0\tilde{\psi}^++\tilde{\alpha}^0\tilde{\psi}^+}{1-\tilde{\phi}^+\tilde{\phi}^-}\ ,\ \ \tilde{\bar{\beta}}(\lambda,\xi)=\frac{\tilde{\phi}^+\tilde{\alpha}^0\tilde{\psi}^-+\tilde{\beta}^0\tilde{\psi}^-}{1-\tilde{\phi}^-\tilde{\phi}^+}\ . \label{eq: expression for alpha nad beta}
\end{equation}

Note that if we substitute the $\tilde{j}_\alpha$ from the expression for $\tilde{\bar{\beta}}$ in~\eqref{eq: beta tilde bar} into $\tilde{j}_\beta=\tilde{\phi}^-\tilde{j}_\alpha+\tilde{\beta}^0\tilde{\phi}^-$ we obtain the relation~\eqref{eq: memory kernel definition} in Section~\ref{sec: macroscopic equation for the synchronised movement}.

\paragraph{Fractional scaling} We consider the following scaling
\begin{equation}
    (t_n,k_n,x_n)\mapsto (t/\e^\theta,\ k/\e^\kappa,\ x/\e^\nu)\ ,
\end{equation}
where $\theta,\ \kappa,\ \nu >0$. We introduce the scaling in the expressions~\eqref{eq: psi} and~\eqref{eq: switching rate}
\begin{equation}
\psi_\e(k)=\Bigl( \frac{\e^\kappa k_0}{\e^\kappa k_0+k}\Bigr)^\mu\ ,\ \ \ \phi_\e(k)=\frac{\mu(\e^\kappa k_0)^\mu}{(\e^\kappa k_0+k)^{\mu+1}}\ ,\ \ \ p_\e(k)=\frac{\mu \e^\kappa}{\e^\kappa k_0+k}\ ,
\end{equation}
and from now on, we take $a=\e^\kappa k_0$.

Now consider the case when the cell  starts to move to the right at $t=0$ from the point $x=0$, then ${\alpha}^0=\varepsilon^v\delta(x)$ where $v>0$ is a constant and ${\beta}^0(x)=0$. 
Since $\tilde{\rho}=\tilde{\bar{\alpha}}+\tilde{\bar{\beta}}$ we have, in the Fourier-Laplace space, using~\eqref{eq: alpha and beta important},
\begin{equation}
    \tilde{\psi}^+_\varepsilon\,\tilde{\psi}_\varepsilon^-\,(\tilde{\rho}-\varepsilon^v)\,=\,\tilde{\psi}_\varepsilon^+\,\tilde{\phi}^-\,\tilde{\bar{\beta}}\,\tilde{\psi}_\varepsilon^+\,+\,\tilde{\psi}_\varepsilon^-\,\tilde{\phi}_\varepsilon^+\,\tilde{\bar{\alpha}}\,\tilde{\psi}_\varepsilon^-\ .\label{eq: fourier-laplace for rho}
\end{equation}
Using the expansions~\eqref{eq: important expansions} and following the steps in Appendix \ref{sec: escaping and arrival rates}-II the above expression can be written as
\begin{equation}
    \varepsilon^v+a\lambda_-\Bigl(\frac{\tilde{\bar{\beta}}}{1-\mu}+\frac{\tilde{\bar{\alpha}}}{2-\mu} \Bigr)+a\lambda_+\Bigl(\frac{\tilde{\bar{\beta}}}{2-\mu}+\frac{\tilde{\bar{\alpha}}}{1-\mu} \Bigr)=a^{\mu-1}\Gamma(-\mu+1)(1-\mu)(\lambda^{\mu-1}_+\tilde{\bar{\beta}}+\lambda^{\mu-1}_-\tilde{\bar{\alpha}})\ .\label{eq: fourier laplace macroscopic equation}
\end{equation}
Replacing $\lambda_\pm=\lambda\pm iv\xi$ and including the scaling we have
\begin{align}
    \varepsilon^v&+k_0(\varepsilon^{\theta+\kappa}\lambda-\varepsilon^{\nu+\kappa}iv\xi)\Bigl(\frac{\tilde{\bar{\beta}}}{1-\mu}+\frac{\tilde{\bar{\alpha}}}{2-\mu} \Bigr)+k_0(\varepsilon^{\theta+\kappa}\lambda+\varepsilon^{\nu+\kappa}iv\xi)\Bigl(\frac{\tilde{\bar{\beta}}}{2-\mu}+\frac{\tilde{\bar{\alpha}}}{1-\mu} \Bigr)\nonumber\\&=k_0^{\mu-1}\varepsilon^{(\kappa+\nu)(\mu-1)}\Gamma(-\mu+1)(1-\mu)\Bigl((iv\xi)^{\mu-1}\tilde{\bar{\beta}}+(-iv\xi)^{\mu-1}\tilde{\bar{\alpha}} \Bigr)\ .
\end{align}
Note that on the right hand side we have used a quasi-static approximation $(\varepsilon^\theta\lambda\pm \varepsilon^\nu iv\xi)^{\mu-1}\simeq \varepsilon^{\nu(\mu-1)}(\pm iv\xi)^{\mu-1}$, assuming $\theta>\nu$.
Grouping terms and using the definitions for the macroscopic density and the local flux
$
\tilde{\rho}=\tilde{\bar{\alpha}}+\tilde{\bar{\beta}}\ , \, \tilde{J}_\e=\frac{\tilde{\bar{\alpha}}-\tilde{\bar{\beta}}}{\e}\ 
$ respectively,
we obtain
\begin{equation}
    \e^{\theta+\kappa}\lambda\tilde{\rho}-c_\mu\e^v+\frac{1}{(3-2\mu)}\e^{\nu+\kappa+1}iv\xi\tilde{J}_\e=d_\mu\e^{(\kappa+\nu)(\mu-1)}\Bigl((iv\xi)^{\mu-1}\tilde{\bar{\beta}}+(-iv\xi)^{\mu-1}\tilde{\bar{\alpha}} \Bigr)\ ,
\end{equation}
where \begin{equation}c_\mu=-\frac{(1-\mu)(2-\mu)}{k_0(3-2\mu)}>0\ , \quad d_\mu=k_0^{\mu-2}\frac{(1-\mu)^2(2-\mu)}{(3-2\mu)}>0\quad \textnormal{for}\quad 1<\mu<3/2\ .\label{eq: values of mu superdiffusion}\end{equation}
Choosing $\theta=\kappa(\mu-2)+\nu(\mu-1)$, $v=(\kappa+\nu)(\mu-1)$ and noting that $\nu+\kappa+1>(\kappa+\nu)(\mu-1)$ (which means that the normal diffusion is of lower order) for $\mu<2+\frac{1}{\kappa+\nu}$ we get
\begin{align}
\lambda\tilde{\rho}-{c_\mu}=d_\mu\Bigl((iv\xi)^{\mu-1}\tilde{\bar{\beta}}+(-iv\xi)^{\mu-1}\tilde{\bar{\alpha}} \Bigr)\ .
\end{align}
 Using
$
\mathcal{FL}[ \partial_t\rho(t,x)]=\lambda\tilde{\rho}(\lambda,\xi)-\rho^0(0)\ ,
$
where we assume that $c_\mu=\rho^0(0)$ and the following relations for $s\in(0,1)$ \cite{ferrari2017some,ferrari2018weyl}
\begin{equation*}
    \mathcal{F}[\mathds{D}^{s}_{-}f]=(i\xi)^s\tilde{f}\ ,\ \ \ \mathcal{F}[\mathds{D}^{s}_+f]=(-i\xi)^s\tilde{f}\ ,
\end{equation*}
we have
\begin{equation*}
    \partial_t\rho=d_\mu(\mathds{D}_-^{\mu-1}{\bar{\beta}}+\mathds{D}_+^{\mu-1}{\bar{\alpha}})\qquad\textnormal{for}\quad \mu\in(1,\ 3/2)\ .
\end{equation*}
Here $\mathds{D}^{s}_{-}$ and $\mathds{D}^{s}_+$ are Riemann-Liouville fractional derivatives defined as
\begin{align*}
\mathds{D}^{s}_{-}f=\frac{-1}{\Gamma(1-s)}\frac{\partial}{\partial x}\int_x^\infty\frac{f(y)}{(y-x)^{s}}\diff y\ ,\quad
\mathds{D}^{s}_+f=\frac{1}{\Gamma(1-s)}\frac{\partial}{\partial x}\int_{-\infty}^x\frac{f(y)}{(x-y)^{s}}\diff y\ .
\end{align*}

 Following the steps in Appendix \ref{sec: escaping and arrival rates}-III we finally write the macroscopic equation as 
\begin{equation}
    \partial_t\rho(t,x)={C}\Bigl(-\frac{\diff^2}{\diff x^2} \Bigr)^{\frac{\mu-1}{2}}\rho(t,x)\ ,
\end{equation}
where
$
{C}={d_\mu}\frac{\mu-1}{2\Gamma(2-\mu)}\frac{1}{c(1,\frac{\mu-1}{2})}>0\ .
$

\section{Numerical results}\label{sec: numerics}

We present some numerical results for the discrete synchronised system where we show the diffusive and superdiffusive regimes, in agreement with the results in Sections \ref{sec: syncrhonisation movement} and \ref{sec: macroscopic equation for the synchronised movement}.

We start with the discrete description of the synchronised movement, which leads, in the limit, to~\eqref{eq: alpha1}-\eqref{eq: beta initial}.

\subsection{Discrete description of the fully synchronised cell movement }
\label{sec: sync movement}

The system~\eqref{eq: alpha1}-\eqref{eq: beta initial} can be derived from a point particle when the probability of moving depends on previous steps taken in the same direction. 
We only treat the full synchronisation case, this derivation can be extended to the non-synchronised system. 

As before, we denote by $\alpha(N,x,k)$ the probability that the cell moves to the right. Here $N$ is the total number of steps, $k$ are the number of steps given by the cell in the same direction, and $x$ is the position. Analogously, we denote by $\beta(N,x,k)$  the probability that the cell moves to the left. 
We recall that the probability of changing the  direction is denoted by $q_k=\tau p_k$ where $\tau$ is a small time step. Therefore the probability of keep moving in the same direction is $\tilde{q}_k=1-\tau p_k$. Since the cell has ``memory'' of the direction of the previous steps, we assume that the probability of changing direction decreases with the number of steps $k$ according to a power-law. This models the directional persistence observed in experiments in \cite{huda2018levy}.

At each time a particle makes a step to the left or to the  right according to its status, and then decides to keep moving in the same direction or reverse direction. 
\\

{
\paragraph{Discrete jumping} We first consider a cell moving to the right,  after $N$ steps, where it gave $k$ steps in this direction.  In the previous step $N-1$, this  cell had done $k-1$ steps to the right and thus the probability of keep moving to the right is 
}
\begin{equation}
\alpha(N,x,k)=(1-\tau p_k)\alpha(N-1,x-\delta,k-1)\ .\label{eq: discrete persistence}
\end{equation}
{
We also have to consider the events when the cell was moving to the left at step $N-1$, described by $\beta(N-1,x,k)$ and reverses direction with probability $\tau p_k$. Since the particle changed direction, it is  set at $k=0$ moving to the right, and thus we have
}
\begin{equation}
    \alpha(N,x,0)=\tau  \sum_{k=1}^{N-1} p_k\beta(N-1,x-\delta,k)\ .\label{eq: reversing direction}
\end{equation}
From~\eqref{eq: discrete persistence} we can write, after dividing by $\tau$
\begin{equation}
\frac{\alpha(N,x,k)-\alpha(N-1,x-\delta,k-1)}{\tau}=-p_k\alpha(N-1,x {-} \delta,k-1)\ .
\end{equation}
In the limit, for $\tau, \delta \to 0$ and $v=\delta/\tau$ we get, 
\begin{equation}
\begin{aligned}
    \partial_t\alpha(t,x,k)+\partial_k\alpha(t,x,k)+v\partial_x\alpha(t,x,k)&=-p(k)\alpha(t,x,k)\ ,\\
    \alpha(t,x,0)&=\int_0^{\infty} p(k)\beta(t,x,k)\diff k\ .
\end{aligned}
\end{equation}
The second relation in~\eqref{eq: alpha1} is obtained from~\eqref{eq: reversing direction}, in the limit. 

Following the same steps for the left movement of the particle we start from
\begin{align}
    \beta(N,x,k)&=(1-\tau p_k)\beta(N-1,x+\delta,k-1)\ ,\nonumber\\ 
    \beta(N,x,0)&=\tau \sum_{k=1}^{N-1} p_k\alpha (N-1, x{ + \delta},k )\ ,
\end{align}
and in the limit we obtain
\begin{equation}
\begin{aligned}
\partial_t\beta(t,x,k)+\partial_k\beta(t,x,k)-v\partial_x\beta(t,x,k)&=-p(k)\beta(t,x,k)\ ,\\
\beta(t,x,0)&=\int_0^{\infty} p(k)\alpha(t,x,k)\diff k\ .\\
\end{aligned}
\end{equation}

\subsection{Numerical set up and main numerical results}

We consider a discrete velocity jump model which describes the left and right movement as in Section~\ref{sec: sync movement}, in an infinite one dimensional domain. We assume that the speed of the cell is constant given by $v=\pm 1$ and the probability of changing direction from left to right is governed by~\eqref{eq: psi}. To decide whether the cell changes direction or not, we 
{ use the rejection method.}
We randomly generate a number between $(0,1)$, if that number is bigger than a probability $P=\frac{\psi(k)}{\psi(k-1)}$\footnote{$\psi(k)$ is the probability of a run of length at least $k$. {I would like to
achieve this distribution by independent decisions whether to turn or
not (based on $\textnormal{rand}(1)$)}. The probability to continue the run after the first time step is $P(1)$, the probability to continue after the second
time step is $P(2)$, etc. The probability that the cell has not
turned within the first $k$ time steps is $P(1)P(2)...P(k)$. This is in fact equal to $\psi(k)$. The formula $\psi(k) = P(1)P(2)...P(k)$ $\forall k$ has a
unique solution for the probabilities $P$: $P(j) = \psi(j)/\psi(j-1)$.
}, 
then the cell changes direction, otherwise it keeps moving without changing. The steps $k$ are updated in each iteration and therefore $P$, where we always start with $k=1$.  The cell updates its position according to $x(t_{i+1})=x(t_i)+v$. This same description can be extended for the non-synchronisation case, where the movement of the front ($y$) and the back ($x$) are independent. 
Every time the cell changes direction we count the number of steps $k$ given in the same direction. For the non-synchronisation case we take into account the  biologically relevant switching probabilities given in Section \ref{subsec: biologically relevant switching prob} to preserve the realistic cell length.

With this toy example we are able to compute the mean square displacement (MSD) $\langle x^2\rangle$ of the cells. As stated in the Introduction, normal diffusion processes are characterised by $\langle x^2 \rangle\sim t$, while for the case of superdiffusion $\langle x^2\rangle\sim t^\zeta$ for $\zeta\in (1,2)$, where $\zeta=\mu/2$.

\begin{figure}[tbhp]
    \hspace{-1cm}\subfloat[]{\label{fig: MSD11}\includegraphics[scale=0.5]{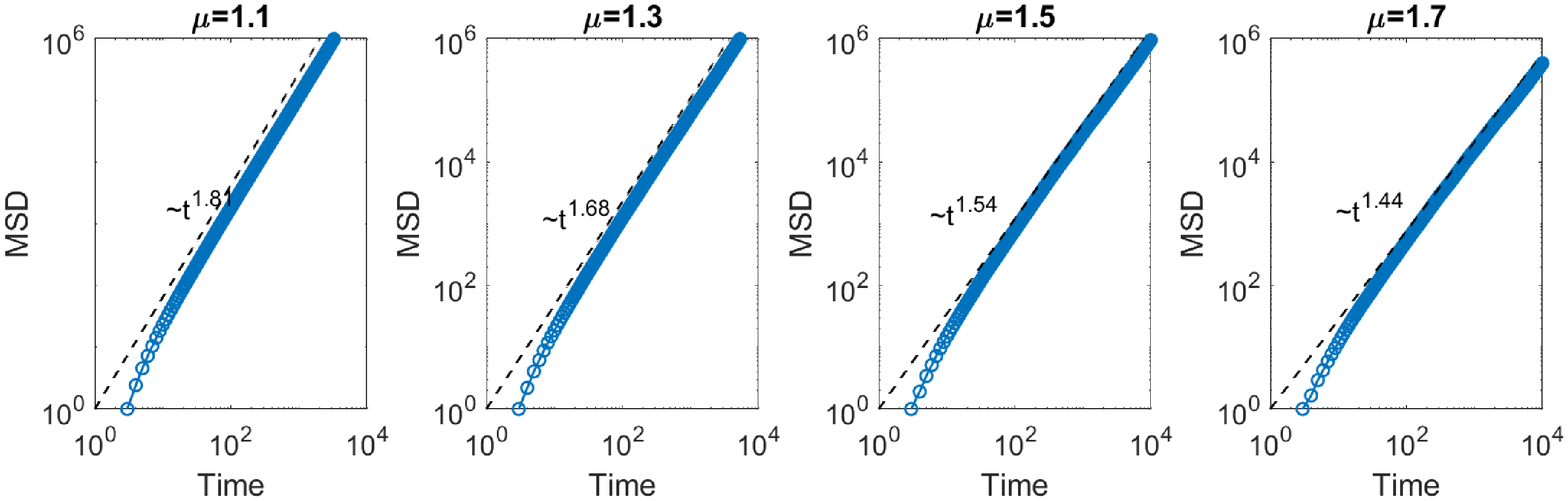}}
    
   \hspace{-1cm}\subfloat[]{\label{fig: MSD15}\includegraphics[scale=0.5]{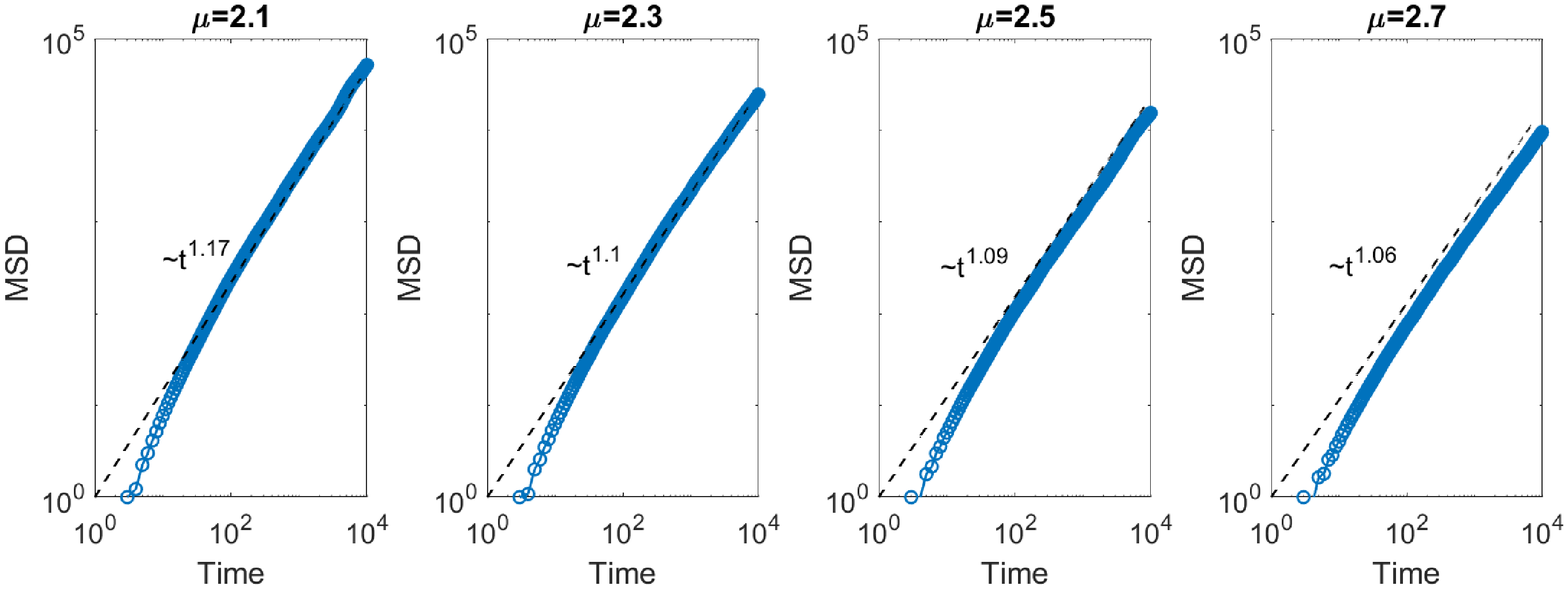}} 
  \caption{Average of the MSD taken from 10000 individual trajectories when we let them run for $t=10000$. (a) describes the superdiffusive regime while  (b) gives the normal diffusion case.}
    \label{fig:sync superdiffusion}
\end{figure}

 In Figure \ref{fig:sync superdiffusion} we have the average of the MSD where this average is taken over $10\ 000$ runs and the trajectories of the cell follows the discrete velocity jump process described before. As obtained in \eqref{eq: values of mu superdiffusion}, the superdiffusion movement for the synchronised case is observed when $\mu\in(1,3/2)$ which agrees with the results in Figure~\ref{fig: MSD11}. 
On the other hand, we consider the normal diffusion limit of the synchronised system derived in Section \ref{sec: diffusion} where we observed normal diffusion for $\mu\in(2,3)$. From Figure \ref{fig: MSD15}, we see that the slope of the MSD is approximately $1$, corresponding to the normal diffusion case.

Moreover, these findings are in agreement with  \cite{huda2018levy}, where the authors observed superdiffusion for L\'{e}vy exponents ${\mu}=1.39,\ 1.58,\ 1.50$ and normal diffusion for $\mu=2.17,\ 2.36,\ 3.57$ (see Table 1 in \cite{huda2018levy}).

\begin{figure}[tbhp]
    \centering
    \includegraphics[scale=0.5]{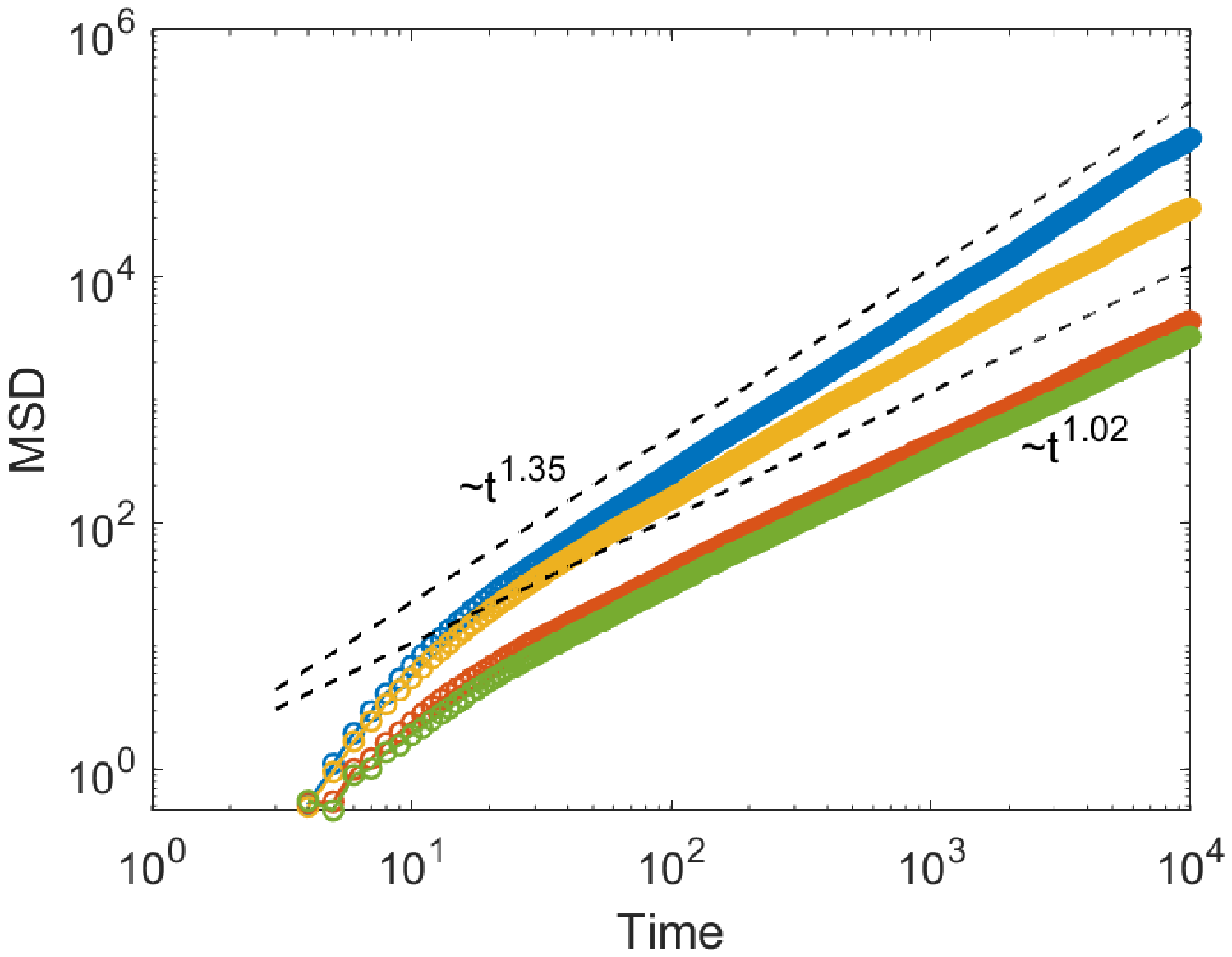}
    \caption{MSD for the non-synchronised case. The blue and yellow lines are for $\mu=1.3$ and $1.5$ respectively, and the orange and green are for the cases $\mu=2.3$ and $2.5$.}
    \label{fig: nonsynchronised case}
\end{figure}

{Finally, for completeness we also present the numerical results for the non-synchronised case in Figure \ref{fig: nonsynchronised case}. Here we observe a similar behaviour as for the synchronised with the difference that now the superdiffusion is ``weaker'' in the sense that even for very small values of $\mu$ the slope of the MSD is close to one. }

\section{Conclusion and perspectives}

We developed a formalism allowing to take into account how eukariotic cells move by protrusions (front of the cell) and retractions (back of the cell), keeping the simplicity of one space dimension for motion. Full generality, assuming that back and front are independent leads to a mathematical model hardly amenable to analysis, but various synchronisation levels lead to simpler models for which macroscopic effects can be observed. Among them we found normal drift-diffusion but more interestingly, instability can occur and, in the fully synchronised case, fractional diffusion characterised by long jumps. This is in accordance with experimental observations in \cite{huda2018levy} where the trajectories of metastatic cells, which move in a synchronised way, followed a power-law distribution, characteristic of a superdiffusion process. 

From a modelling and analytical point of view, several questions are left open. For instance, a better understanding of the full model and of possible model reduction. Also, the introduction of more biological details, for example, in the switching direction probability \eqref{eq: psi}. We could tailor this function to a specific system by knowing the internal mechanisms that leads to synchronisation in cells. Moreover, we could extend our model to several dimensions and connect it to models of cell polarisation such as \cite{loy2020kinetic,loy2020modelling}. Finally, it would be interesting to look at the effect of the interactions with the environment and collective effects.



\appendix

\section{Miscellaneous}
\label{sec: escaping and arrival rates}

(I) We compute the escape and arrival rates introduced in~\eqref{eq: switching rates j} by using the characteristic solutions~\eqref{eq: characteristic alpha} and~\eqref{eq: characteristic beta}. We start from
\[
j_\beta(t,x)=\int_0^{t}p(k)\beta(t,x,k)\diff x
\]
which, by using~\eqref{eq: switching rate} and~\eqref{eq: characteristic beta}, can be re-written as
\begin{align}
    j_\beta(t,x) & =\int_0^{t}\phi(k)\beta(t-k,x+vk,0)\diff k+{\bar{\beta}^0(x+vk)\phi(k)}\nonumber\\
    &= \int_0^{t} \phi(t-s)e^{v(t-s)\partial_x}\beta(s,x,0)\diff s \, +{\bar{\beta}^0(x+vk)\phi(k)}\ .
\end{align}
The last equality is obtained using the change of variables $k=t-s$ along with the following Taylor expansion
\begin{align}
    e^{v(t-s)\partial_x}f(x)&=\sum_{m=0}^\infty\frac{(v(t-s)\partial_x)^m}{m!}f(x)\nonumber\\ & =\sum_{m=0}^\infty\frac{1}{m!}(v(t-s))^m\partial_x^mf(x)=f(x+v(t-s))\ .\nonumber
\end{align}
Analogously we can obtain \eqref{eq: def j_alpha} for $j_\alpha(t,x)$.\\

\noindent (II) Now we aim to derive the expression \eqref{eq: fourier-laplace for rho}. From \eqref{eq: alpha and beta important} we write, after multiplying both sides by $\tp^+\tp^-$
\begin{equation}
\tp^+\,\tp^-\,\tilde{\rho}\,=\,\tp^+\,\tph^-\,\tilde{\bar{\beta}}\,+\,\tilde{\alpha}^0\,\tp^+\,\tp^-\,+\,\tp^-\,\tph^+\,\tilde{\bar{\alpha}}\,\tp^-\,+\,\tilde{\beta}^0\,\tp^-\,\tp^+\ .\label{eq: relation rho appendix}
\end{equation}
Using the initial conditions ${\alpha}_\varepsilon^0=\varepsilon^z\delta(x)$ and ${\beta}^0(x)=0$ we obtain \eqref{eq: fourier-laplace for rho}. Now, we introduce the scaling to \eqref{eq: important expansions} and we write
\begin{align*}
    \tp^\pm&=-\frac{a}{1-\mu}-\frac{a^2\lambda_\pm}{(1-\mu)(2-\mu)}+a^\alpha\lambda_\pm^{\mu-1}\Gamma(-\mu+1)+\mathcal{O}(a^3\lambda_\pm^2)\ ,\\
    \tph^\pm&=1+\frac{a\lambda_\pm}{1-\mu}+a^\mu\lambda_\pm^\mu+O(\lambda^{\mu+1}) \ .
\end{align*}
Hence from here we compute
\begin{align*}
    \tp^+\tph^-&=-\frac{a}{1-\mu}-\frac{a^2\lambda_+}{(1-\mu)(2-\mu)}+a^\mu\lambda_+^{\mu-1}\Gamma(-\mu+1)-\frac{a^2\lambda_-}{(1-\mu)^2}+\mathcal{O}(a^{\mu+1})\ ,\\
    \tp^-\tph^+&=-\frac{a}{1-\mu}-\frac{a^2\lambda_-}{(1-\mu)(2-\mu)}+a^\mu\lambda_-^{\mu-1}\Gamma(-\mu+1)-\frac{a^2\lambda_+}{(1-\mu)^2}+\mathcal{O}(a^{\mu+1})\ ,\\
    \tp^+\tp^-&=\frac{a^2}{(1-\mu)^2}-a^{\mu+1}\frac{\Gamma(\mu+1)}{1-\mu}(\lambda_-^{\mu-1}+\lambda^{\mu-1}_+)+\mathcal{O}(a^3)\ .
\end{align*}
Substituting these three quantities in \eqref{eq: relation rho appendix} we arrive at \eqref{eq: fourier laplace macroscopic equation}.\\

\noindent(III) Finally, we are going to work only with the fractional operators. Following \cite{ferrari2017some,ferrari2018weyl} we have
\begin{align}\mathds{D}_-^{\mu-1}\bar{\beta}&=\frac{-1}{\Gamma(2-\mu)}\frac{\partial}{\partial x}\int_x^\infty\frac{\bar{\beta}(s)}{(s-x)^{\mu-1}}\diff s=\frac{\mu-1}{\Gamma(2-\mu)}\int_0^\infty\frac{\bar{\beta}(x)-\bar{\beta}(x+s)}{s^{\mu}}\diff s\nonumber\\
\mathds{D}_+^{\mu-1}\bar{\alpha}&=\frac{1}{\Gamma(2-\mu)}\frac{\partial}{\partial x}\int^x_{-\infty}\frac{\bar{\alpha}(s)}{(s-x)^{\mu-1}}\diff s=\frac{\mu-1}{\Gamma(2-\mu)}\int_0^\infty\frac{\bar{\alpha}(x)-\bar{\alpha}(x-s)}{s^{\mu}}\diff s\nonumber\ .
\end{align}
The above relation is true if $\bar{\alpha},\bar{\beta}\in C^1(\mathds{R})$ and $\bar{\alpha},\bar{\beta}=o(|x|^{\mu-2-\epsilon})$, $x\to +\infty$ for $\epsilon>0$ (equivalence between Marchaud derivative and Riemann-Liuoville derivative).

Now we are going to use the fact that the sum $\mathds{D}_-^{\mu-1}f+\mathds{D}^{\mu-1}_+f$ gives the fractional Laplace operator in one dimension, also known as the Riesz derivative,
\begin{align*}
    \mathds{D}_-^{\mu-1}\bar{\beta}+\mathds{D}_+^{\mu-1}\bar{\alpha}&=\frac{\mu-1}{\Gamma(2-\mu)}\Bigl(\int_0^\infty \frac{\bar{\beta}(x)-\bar{\beta}(x+s)}{s^\mu}\diff s+\int_0^\infty\frac{\bar{\alpha}(x)-\bar{\alpha}(x-s)}{s^\mu}\diff s \Bigr)\\
    &=\frac{\mu-1}{\Gamma(2-\mu)}\Bigl(\int_{-\infty}^0 \frac{\bar{\beta}(x)-\bar{\beta}(x-s)}{|s|^\mu}\diff s+\int_0^\infty\frac{\bar{\alpha}(x)-\bar{\alpha}(x-s)}{s^\mu}\diff s \Bigr)\\
    &=\frac{\mu-1}{2\Gamma(2-\mu)}\int_{-\infty}^\infty\frac{\bar{\beta}(x)+\bar{\alpha}(x)-\bar{\beta}(x-s)-\bar{\alpha}(x-s)}{|s|^\mu}\diff s\\
    &=\frac{\mu-1}{2\Gamma(2-\mu)}\int_{-\infty}^\infty\frac{\rho(x)-\rho(x-s)}{|s|^\mu}\diff s=\frac{\mu-1}{2\Gamma(2-\mu)}\frac{1}{c(1,\frac{\mu-1}{2})}\Bigl(-\frac{\diff^2}{\diff x^2} \Bigr)^{\frac{\mu-1}{2}}\rho(t,x)\ ,
\end{align*}
where $c(1,\frac{\mu-1}{2})$ is a normalization constant.

\clearpage

\bibliographystyle{abbrv}
\bibliography{persistent_walk}

\begin{thebibliography}{10}

\bibitem{ariel2015swarming}
G.~Ariel, A.~Rabani, S.~Benisty, J.~D. Partridge, R.~M. Harshey, and A.~Be'Er.
\newblock Swarming bacteria migrate by {L}{\'e}vy walk.
\newblock {\em Nature Communications}, 6(1):1--6, 2015.

\bibitem{NIST:DLMF}
{\it NIST Digital Library of Mathematical Functions}.
\newblock http://dlmf.nist.gov/, Release 1.0.14 of 2016-12-21.
\newblock F.~W.~J. Olver, A.~B. {Olde Daalhuis}, D.~W. Lozier, B.~I. Schneider,
  R.~F. Boisvert, C.~W. Clark, B.~R. Miller and B.~V. Saunders, eds.

\bibitem{estrada2018fractional}
G.~Estrada-Rodriguez, H.~Gimperlein, and K.~J. Painter.
\newblock Fractional {P}atlak--{K}eller--{S}egel equations for chemotactic
  superdiffusion.
\newblock {\em SIAM Journal on Applied Mathematics}, 78(2):1155--1173, 2018.

\bibitem{fedotov2015persistent}
S.~Fedotov, A.~Tan, and A.~Zubarev.
\newblock Persistent random walk of cells involving anomalous effects and
  random death.
\newblock {\em Physical Review E}, 91(4):042124, 2015.

\bibitem{ferrari2017some}
F.~Ferrari.
\newblock Some nonlocal operators in the first {H}eisenberg group.
\newblock {\em Fractal and Fractional}, 1(1):15, 2017.

\bibitem{ferrari2018weyl}
F.~Ferrari.
\newblock Weyl and {M}archaud derivatives: A forgotten history.
\newblock {\em Mathematics}, 6(1):6, 2018.

\bibitem{focardi2009adaptive}
S.~Focardi, P.~Montanaro, and E.~Pecchioli.
\newblock Adaptive {L}{\'e}vy walks in foraging fallow deer.
\newblock {\em PLoS One}, 4(8):e6587, 2009.

\bibitem{frankgoudon}
M.~Frank and T.~Goudon.
\newblock On a generalized {B}oltzmann equation for non-classical particle
  transport.
\newblock {\em Kinetic and Related Models}, 3(3):395--407, 2010.

\bibitem{franksun2018}
M.~Frank and W.~Sun.
\newblock Fractional diffusion limits of non-classical transport equations.
\newblock {\em Kinetic and Related Models}, 11(6):1503--1526, 2018.

\bibitem{fricke2016persistence}
G.~M. Fricke, K.~A. Letendre, M.~E. Moses, and J.~L. Cannon.
\newblock Persistence and adaptation in immunity: T cells balance the extent
  and thoroughness of search.
\newblock {\em PLoS Computational Biology}, 12(3):e1004818, 2016.

\bibitem{harrisgeneralized}
T.~H. Harris, E.~J. Banigan, D.~A. Christian, C.~Konradt, E.~D.~T. Wojno,
  K.~Norose, E.~H. Wilson, B.~John, W.~Weninger, A.~D. Luster, et~al.
\newblock Generalized l{\'e}vy walks and the role of chemokines in migration of
  effector cd8+ t cells.
\newblock {\em Nature}, 486(7404):545--548, 2012.

\bibitem{huda2018levy}
S.~Huda, B.~Weigelin, K.~Wolf, K.~V. Tretiakov, K.~Polev, G.~Wilk, M.~Iwasa,
  F.~S. Emami, J.~W. Narojczyk, M.~Banaszak, et~al.
\newblock L{\'e}vy-like movement patterns of metastatic cancer cells revealed
  in microfabricated systems and implicated in vivo.
\newblock {\em Nature Communications}, 9(1):1--11, 2018.

\bibitem{Olla2009}
M.~Jara, T.~Komorowski, and S.~Olla.
\newblock Limit theorems for additive functionals of a {M}arkov chain.
\newblock {\em The Annals of Applied Probability}, 19(6):2270--2300, 2009.

\bibitem{korobkova2004molecular}
E.~Korobkova, T.~Emonet, J.~M. Vilar, T.~S. Shimizu, and P.~Cluzel.
\newblock From molecular noise to behavioural variability in a single
  bacterium.
\newblock {\em Nature}, 428(6982):574--578, 2004.

\bibitem{li2008persistent}
L.~Li, S.~F. N{\o}rrelykke, and E.~C. Cox.
\newblock Persistent cell motion in the absence of external signals: a search
  strategy for eukaryotic cells.
\newblock {\em PLoS One}, 3(5):e2093, 2008.

\bibitem{loy2020kinetic}
N.~Loy and L.~Preziosi.
\newblock Kinetic models with non-local sensing determining cell polarization
  and speed according to independent cues.
\newblock {\em Journal of Mathematical Biology}, 80(1):373--421, 2020.

\bibitem{loy2020modelling}
N.~Loy and L.~Preziosi.
\newblock Modelling physical limits of migration by a kinetic model with
  non-local sensing.
\newblock {\em Journal of Mathematical Biology}, 80(6):1759--1801, 2020.

\bibitem{M3}
A.~Mellet, S.~Mischler, and C.~Mouhot.
\newblock Fractional diffusion limit for collisional kinetic equations.
\newblock {\em Arch. Ration. Mech. Anal.}, 199(2):493--525, 2011.

\bibitem{metzler2000random}
R.~Metzler and J.~Klafter.
\newblock The random walk's guide to anomalous diffusion: a fractional dynamics
  approach.
\newblock {\em Physics Reports}, 339(1):1--77, 2000.

\bibitem{Pmoussa2019}
A.~Moussa, B.~Perthame, and D.~Salort.
\newblock Backward parabolicity, cross-diffusion and {T}uring instability.
\newblock {\em J. Nonlinear Sci.}, 29(1):139--162, 2019.

\bibitem{othmer1988models}
H.~G. Othmer, S.~R. Dunbar, and W.~Alt.
\newblock Models of dispersal in biological systems.
\newblock {\em Journal of Mathematical Biology}, 26(3):263--298, 1988.

\bibitem{Pyasuda2018}
B.~Perthame and S.~Yasuda.
\newblock Stiff-response-induced instability for chemotactic bacteria and
  flux-limited {K}eller-{S}egel equation.
\newblock {\em Nonlinearity}, 31(9):4065--4089, 2018.

\bibitem{raichlen2014evidence}
D.~A. Raichlen, B.~M. Wood, A.~D. Gordon, A.~Z. Mabulla, F.~W. Marlowe, and
  H.~Pontzer.
\newblock Evidence of {L}{\'e}vy walk foraging patterns in human
  hunter--gatherers.
\newblock {\em Proceedings of the National Academy of Sciences},
  111(2):728--733, 2014.

\bibitem{reynolds2018levy}
A.~Reynolds, E.~Ceccon, C.~Baldauf, T.~Karina~Medeiros, and O.~Miramontes.
\newblock L{\'e}vy foraging patterns of rural humans.
\newblock {\em PLoS One}, 13(6):e0199099, 2018.

\bibitem{reynolds2017weierstrassian}
A.~Reynolds, G.~Santini, G.~Chelazzi, and S.~Focardi.
\newblock The weierstrassian movement patterns of snails.
\newblock {\em Royal Society open science}, 4(6):160941, 2017.

\bibitem{reynolds2007displaced}
A.~M. Reynolds, A.~D. Smith, R.~Menzel, U.~Greggers, D.~R. Reynolds, and J.~R.
  Riley.
\newblock Displaced honey bees perform optimal scale-free search flights.
\newblock {\em Ecology}, 88(8):1955--1961, 2007.

\bibitem{sims2008scaling}
D.~W. Sims, E.~J. Southall, N.~E. Humphries, G.~C. Hays, C.~J. Bradshaw, J.~W.
  Pitchford, A.~James, M.~Z. Ahmed, A.~S. Brierley, M.~A. Hindell, et~al.
\newblock Scaling laws of marine predator search behaviour.
\newblock {\em Nature}, 451(7182):1098--1102, 2008.

\bibitem{sokolov2003towards}
I.~M. Sokolov and R.~Metzler.
\newblock Towards deterministic equations for {L}{\'e}vy walks: The fractional
  material derivative.
\newblock {\em Physical Review E}, 67(1):010101, 2003.

\bibitem{viswanathan1996levy}
G.~M. Viswanathan, V.~Afanasyev, S.~V. Buldyrev, E.~Murphy, P.~Prince, and
  H.~E. Stanley.
\newblock L{\'e}vy flight search patterns of wandering albatrosses.
\newblock {\em Nature}, 381(6581):413--415, 1996.

\bibitem{zaburdaev2015levy}
V.~Zaburdaev, S.~Denisov, and J.~Klafter.
\newblock L{\'e}vy walks.
\newblock {\em Reviews of Modern Physics}, 87(2):483, 2015.

\end{thebibliography}
\end{document}